\documentclass[preprint,aps,nofootinbib]{revtex4}

\usepackage{dcolumn}
\usepackage{bm}
\usepackage{epsfig}
\usepackage{graphicx,subfigure}

\usepackage{amsmath}
\usepackage{amssymb}
\usepackage{color}
\usepackage[usenames,dvipsnames]{xcolor}
\usepackage{hyperref}

\def\fun#1#2{\lower3.6pt\vbox{\baselineskip0pt\lineskip.9pt
  \ialign{$\mathsurround=0pt#1\hfil##\hfil$\crcr#2\crcr\sim\crcr}}}

\input epsf

\newcommand{\be}{\begin{equation}}
\newcommand{\ee}{\end{equation}}
\newcommand{\bea}{\begin{eqnarray}}
\newcommand{\eea}{\end{eqnarray}}

\newcommand{\ns}{\color{Orchid}}

\begin{document}

\begin{flushright}
\vspace*{-1.5cm}
\vspace{-0.2cm}ANL-HEP-PR-13-14\\
\vspace{-0.2cm}EFI-13-3 \\
\vspace{-0.2cm}FERMILAB-PUB-13-065-T\\
\vspace{-0.2cm}MCTP-13-07 
\end{flushright}

\vspace*{-0.5cm}

\title{Light Stops, Light Staus and the 125~GeV Higgs}

\author{
\vspace{0.1cm} 
\mbox{\bf Marcela Carena$^{\,a,b,c}$, Stefania Gori$^{\,b,d}$,}\\
\mbox{\bf Nausheen R. Shah$^e$,  Carlos E.~M.~Wagner$^{\,b,c,d}$ and Lian-Tao Wang$^{\,b,c}$ }
 }
\affiliation{
\vspace*{.1cm}
$^a$  \mbox{Fermi National Accelerator Laboratory, P.O. Box 500, Batavia, IL 60510}~\footnote{\href{http://theory.fnal.gov}{http://theory.fnal.gov}} \\
$^b$  \mbox{Enrico Fermi Institute, University of Chicago, Chicago, IL 60637}\\
$^c$  \mbox{Kavli Institute for Cosmological Physics, University of Chicago, Chicago, IL 60637}\\
$^d$ \mbox{High Energy Physics Division, Argonne National Laboratory, Argonne, IL 60439}\\
$^e$ \mbox{Michigan Center for Theoretical Physics, Department of Physics,} \\
\mbox{University of Michigan, Ann Arbor, MI 48109}\\
}

\begin{abstract}
The ATLAS and CMS experiments have recently announced the discovery of a Higgs-like resonance with mass close to 125~GeV. Overall, the data is consistent with a Standard Model (SM)-like Higgs boson. Such a particle may arise in the minimal supersymmetric extension of the SM with average stop masses of the order of the TeV scale and a sizable stop mixing parameter.  In this article we discuss properties of the SM-like Higgs production and decay rates induced by the possible presence of light staus and light stops. Light staus can affect the decay rate of the Higgs into di-photons and, in the case of sizable left-right mixing,  induce an enhancement in this  production channel up to $\sim$~50\% of the Standard Model rate. Light stops may induce sizable modifications of the Higgs gluon fusion production rate and correlated modifications to the Higgs diphoton decay. Departures from SM values of the bottom-quark and tau-lepton couplings to the Higgs can be obtained due to Higgs mixing effects triggered by light third generation scalar superpartners. We describe the phenomenological implications of light staus on searches for light stops and non-standard Higgs bosons.  Finally, we discuss the current status of the search for light  staus produced in association with sneutrinos, in final states containing a $W$ gauge boson and a pair of $\tau$s. 
\end{abstract}

\maketitle

\section{Introduction}

The ATLAS and CMS experiments have recently announced the discovery of a new bosonic resonance with mass close to 125~GeV~\cite{:2012gk,:2012gu}. The production and decay rates of this new particle are roughly consistent with those of the Standard Model~(SM) Higgs. Therefore, it is natural to assume that it is indeed a Higgs boson, with similar but not necessarily identical properties as the SM Higgs boson. Hence, its properties should be precisely studied. In particular, 
deviations of its production and decay rates from the SM values may provide the first evidence of new physics at the weak scale. 

Although, current data shows no statistically significant deviation of the signal from the SM predictions, a small enhancement of the diphoton production rate has been observed at ATLAS. This enhancement is present both in the zero and one extra jet channels~(dominated by gluon fusion Higgs production), as well as in the dijet channel~(dominated by weak boson fusion production). The deviation of the Higgs diphoton rate with respect to the SM expectation is somewhat larger than 2-$\sigma$ at ATLAS~\cite{:2012gk,:2012gu,Diphoton}. On the contrary, the CMS analysis of the full data set does not show a similar enhancement in the diphoton rate. Though the early 7 and 8 TeV data hinted towards a small excess of events above the SM prediction, the newest analysis suggests that the Higgs diphoton rate is somewhat suppressed but  within 1-$\sigma$ of the SM expectation at CMS~\cite{CMSdiphoton}

The apparent deviation of the diphoton production rate from the SM predictions has led many authors to investigate the possibility of having an enhancement of the rate of the Higgs decaying to diphotons through charged particle loops~\cite{Carena:2011aa,Carena:2012gp,Blum:2012ii,SchmidtHoberg:2012yy,Kniehl:1995tn,CLW,Batell:2011pz}, through the mixing of the Higgs with other scalar states~\cite{Hall:2011aa}, or both~\cite{Carena:2011aa,Batell:2012mj}. Currently, the  rate of the Higgs-induced $ZZ$ and $WW$ production channels analyzed at both experiments do not present any clear deviation from the SM ones. Results are within about 1-$\sigma$ of the SM expectation, albeit with somewhat large errors~\cite{Dibosons}. A measurement of the $\tau^+\tau^-$ decay rate of the Higgs produced in vector boson fusion has been reported at both ATLAS and CMS, and seems to also be consistent with the SM values within 1-$\sigma$~\cite{LHCtautau}.  Additionally, a search for the associated production of the Higgs with weak gauge bosons, with the Higgs decaying into $b\bar{b}$, has been performed at both the Tevatron~\cite{Aaltonen:2012qt} and the LHC experiments~\cite{Chatrchyan:2012ww,Aad:2012gxa}. Again, the rates are consistent with those expected in the SM~\footnote{However, evidence for Higgs decaying to $b\bar{b}$ has not been observed at ATLAS.}.

In previous works we  have discussed the possible  modification of the diphoton rate via one-loop corrections induced by the  presence of light, highly mixed stau, as well as by a suppression of the  Higgs to $b\bar{b}$ decay rate induced at the one-loop level~\cite{Carena:2011aa,Carena:2012gp}. However, the possible modifications to the gluon fusion production rate,  and to the ratio of $\Gamma(h \to b\bar{b})/\Gamma(h \to \tau^+\tau^-)$ have not been discussed in detail in this framework. 

Modification of the gluon fusion production cross section may be achieved through light stop loops. It is important to note that in the presence of light staus, very light stops ($\sim 100-200$ GeV) may avoid current experimental bounds. This is because the stop decays may get altered, compared to the standard ones considered in current light stop LHC searches. In the presence of light staus and light stops, there may be relevant enhancements or suppressions of the total diphoton production rate, as well as large differences between the Higgs-induced diphoton rates in gluon fusion and vector boson fusion channels.  

The ratio of the $(h \to b\bar{b})$ to $(h \to \tau^+\tau^-)$ Higgs decay widths is important since a departure of its value from the SM one would be clear evidence for new physics~(NP). Moreover, it would also be a clear deviation of the MSSM Higgs sector from type-II two Higgs doublet models (2HDMs). It is the aim of this paper to provide a detailed analysis of these possibilities.  

Overall, assuming no strong violation of  custodial symmetry, the relevant Higgs production and decay rates into SM particles can be parametrized in terms of six effective couplings to: $\{VV$, $\gamma\gamma$, $gg$, $t\bar{t}$, $\tau^+\tau^-$, $b\bar{b}\}$, where all these couplings may deviate from their SM values. In addition, possible Higgs decays into invisible particles may appear, introducing an additional degree of freedom: the Higgs total width (see Ref.~\cite{Dobrescu:2012td} for a recent fit of the current Higgs data). As stressed above, in this article we shall consider  the possible variations of the above couplings in the  presence of light staus and light stops within the Minimal Supersymmetric Standard Model (MSSM).

One may also consider the impact of light staus on heavy Higgs searches and the prospects of detecting light staus at the LHC through their associated production with sneutrinos. This can be analyzed by looking at $(pp\to\tilde\tau^+_1\tilde\nu_\tau)$, with $(2\tau+\ell+\rm{MET})$ final state, where one tau decays leptonically and one hadronically.  This is the same final state as for the search for a Higgs boson decaying into two taus and produced in association with a $W$ boson~\cite{cms_HZtautau}. We will show that indeed this Higgs search may be used to put bounds on the associated production, $(pp\to\tilde\tau^+_1\tilde\nu_\tau)$, once the LHC  accumulates more statistics. 
 
The article is organized as follows. Sec.~\ref{Lss} presents a short review of the  possible effects induced by the presence of light staus on Higgs properties. In Sec.~III we discuss the possible modification of the gluon fusion rate via the existence of relatively light stops. This is followed by a brief discussion of the Tevatron and LHC stop mass bounds  in the presence of light staus. Sec.~IV presents a detailed discussion of possible modifications of the Higgs couplings to bottom quarks and tau leptons. These effects may only be obtained for values of the CP-even Higgs mixing angle which deviate from the ones obtained in the decoupling limit. This implies moderate values of the heavy Higgs masses, leading to possible strong bounds from LHC heavy Higgs searches. The LHC bound on $m_A$ in the presence of light staus are therefore discussed.  In Sec.~V, we present the prospects of detecting a light stau at the LHC in associated production with sneutrinos. We reserve Sec.~VI for our conclusions.

\section{Light Staus and Higgs Decays}
\label{Lss}

One of the simplest possibilities to modify the Higgs to diphoton rate, while leaving all the other Higgs rates SM-like, is the addition of new charged matter particles, with no color and with masses of the order of the weak scale. According to the low energy Higgs theorem~\cite{Ellis:1975ap,Shifman:1979eb} (see also Refs.~\cite{Kniehl:1995tn,CLW}), this may lead to constructive interference with the SM Higgs decay amplitude, if
\begin{equation}
\frac{\partial \log\det\left[M^2(v)\right]}{\partial \log v}\; <\; 0\;,
\end{equation}
where $M^2(v)$ is the mass matrix of the new particles introduced in the loop. Within the MSSM such contributions may come from a light charged Higgs, light charginos, or light sleptons. 

The couplings of the charginos and the charged Higgs to the Higgs are dictated by weak gauge couplings. In addition, their contribution to the diphoton decay amplitude is suppressed for moderate to large values of $\tan\beta$, such as those necessary to obtain a 125~GeV Higgs mass with stops at the TeV scale. Therefore, within the MSSM, charginos lead to at most a correction of the order of $\sim 20\%$ to the SM Higgs diphoton decay width~\cite{Diaz:2004qt,Blum:2012ii}, and the charged Higgs contributions are even smaller~\cite{Blum:2012ii,SchmidtHoberg:2012yy,Altmannshofer:2012ar} (at the level of a few percent).

Concerning possible slepton contributions, we first note that at moderate or large values of $\tan\beta$, the SM-like Higgs is associated with $H_u$, the Higgs that couples to right-handed up-quarks at tree level. The coupling of $H_u$ to sleptons is dominated by the trilinear coupling coming from the  $F$-term contribution, proportional to $(h_{\tau} \mu)$, where $h_{\tau}$ is the $\tau$-Yukawa coupling and $\mu$ is the Higgsino mass parameter. Therefore, a sizable coupling may only be obtained for relatively large values of $\mu$ and large values of $\tan\beta$, which is when the $\tau$-Yukawa coupling is large. Using a normalization in which the sum of the dominant $W$ and top  contributions to the Higgs diphoton decay amplitude in the SM is approximately (-13), for masses larger than or of the order of the Higgs mass, the stau contribution to this amplitude may be approximated by
\begin{equation}
b_{\tilde{\tau}}\; \frac{\partial \log\det\left(M_{\tilde{\tau}}^2\right)}{\partial \log v}  \simeq -\frac{2}{3}~\frac{  m_\tau^2 }{  m^2_{\tilde{\tau}_1} m^2_{\tilde{\tau}_2}}~ \mu^2 \tan^2\beta\, .
\label{amplitude}
\end{equation}
The stau contribution, Eq.~(\ref{amplitude}), needs to be negative and of order one to lead to a relevant enhancement of the diphoton rate. The rate is therefore enhanced for large values of $(\mu \tan\beta)$ and small values of the stau masses. However, for small values of the stau masses and large values of $(\mu \tan\beta)$,  new charge breaking minima are induced and the physical vacuum may become metastable~\cite{Ratz:2008qh,Hisano:2010re, Stability}.  The constraints from vacuum stability place an upper bound on the possible value of $(\mu \tan\beta)$ and hence on the possible loop-induced diphoton rate enhancement~\cite{Kitahara:2012pb,Stability}.  For a given value of $\tan\beta$ this upper bound may be slightly relaxed after considering one-loop corrections to the $\tau$ and $b$ mass~\cite{deltamb,deltamb1,deltamb2,deltamb3}~\footnote{Recently, it has been shown that the bound may also be slightly relaxed by imposing a large hierarchy between the two stau soft masses, $m_{L_3}$ and $m_{e_3}$~\cite{Kitahara:2013lfa}.}. This allows for corrections of up to about $50$ \%~\cite{Stability} for values of the lightest stau mass close to the LEP experimental limit of about 95 GeV~\cite{LEPstau}~\footnote{The stau mass limit drops to values lower than 90~GeV for small values of the neutralino mass or for small differences between the stau and the neutralino masses.}.
For instance, for values of the soft breaking parameters of the order of the weak scale, $m_{L_3} \simeq m_{e_3}\simeq 250$~GeV and  $\tan\beta \simeq 60$, one can obtain  enhancements of the diphoton decay width of order 40\% for values of $\mu$ of order 470~GeV,  small values of $A_{\tau}$ and large values of the CP-odd Higgs mass, $m_A$. Somewhat larger enhancements may be obtained for larger values of $\tan \beta$ ($\sim 70$), without being in conflict with the perturbativity of the Yukawa couplings upto the GUT scale~\cite{Stability}.

In addition to the loop effects induced by light staus, the diphoton rate may be modified by Higgs mixing effects~\cite{Carena:2011aa}. For large values of $\tan\beta$, the loop contributions to the off-diagonal element of the CP-even Higgs mass matrix can efficiently compete with the  $(1/\tan\beta)$ suppressed (but $m_A^2$ enhanced) tree-level value~\cite{Carena:1999bh}. Therefore, the lightest CP-even Higgs boson can have an $H_u$ component even larger than the one obtained in the decoupling limit. This in turn induces a suppression of the bottom quark decay width, and consequently an enhancement of the subdominant decay branching ratios.  In the light stau scenario, this can be achieved for large positive values of the trilinear coupling, $A_{\tau} \simeq 1$~TeV. These effects are in general not expected to be very large since for  values of $\tan\beta\gtrsim 60$, the null LHC results in searches for the heavy Higgs bosons imply that   $m_A\geq 800$ GeV.  In addition, the requirement of vacuum stability severely restricts the large values of $\mu$ and $A_\tau$ for which these effects become relevant~\cite{Stability}.  Therefore, in the MSSM, the effects of Higgs mixing cannot further sizably enhance the Higgs diphoton rate.

These Higgs mixing effects also lead to a suppression of the Higgs decay width into pairs of tau leptons. At tree-level,  the enhancement, or suppression, of the Higgs decay width into bottom or tau pairs with respect to their SM values,  is the same. However, this equality is broken due to the loop-induced couplings of the bottom quarks and tau leptons to the up-type Higgs, $H_u$. We shall discuss these effects in detail in Section \ref{BtauCoup}.  

The Higgs couplings to photons, bottom quarks and tau leptons are modified in a scenario with light staus. However, one needs relatively large values of the CP-odd Higgs mass  to satisfy the LHC constraints, as well as constraints from flavor physics~\cite{Haisch:2012re,Altmannshofer:2012ks}. This in turn implies that the Higgs couplings to the $W$-gauge boson and to the top quark remain close to their SM values. 

Finally,  if we also impose the requirement of a dark matter particle giving rise to the experimentally observed relic density, we are led to the presence of a Bino like lightest neutralino with a mass in the 30-80 GeV range~\cite{Carena:2012gp}. If the neutralino mass is less than $m_h/2$, the Higgs can decay into a pair of lightest neutralinos. However, for large values of $\tan\beta$ and moderate values of $\mu$, the Higgs invisible width is suppressed, and its branching ratio remains of the order of a few percent in the whole region of parameters consistent with the diphoton rate enhancement.

\section{Gluon Fusion and Stop phenomenology}
\label{GluonFusion}

\subsection{Stop Effects in the Higgs Production Cross Section}

In the SM, the gluon fusion amplitude is  predominantly governed by top-quark loops. In the MSSM, there may be relevant contributions coming from the superpartners of the third generation quarks~\cite{Djouadi:1998az,Dermisek:2007fi}. At large values of $\tan\beta$, the modifications can come from both the stop and sbottom sectors. 

As happens with the staus, the most relevant coupling of the Higgs to the sbottoms is proportional to the sbottom mixing parameter, proportional to $(\mu \tan\beta)$. The sbottom contributions  are opposite in sign to the top-quark contribution to the Higgs gluon coupling and lead to a reduced gluon fusion cross section.  However, due to the strong bounds on sbottom masses from the LHC for light neutralinos (lighter than the staus)~\cite{Sbottomlimits} and to the fact that $(\mu \tan\beta)$ is bounded from above by vacuum stability constraints, we find that sbottom loops lead to only minor modifications of the gluon production rate. 

The stop contributions, instead, can be of either sign, depending on the magnitude of the stop mixing parameter, $A_t$, relative to the stop soft masses, as we will discuss in detail below. The relevance of the stop contributions depends strongly on the lightest stop mass, becoming larger for smaller values of $m_{\tilde{t}_1}$. 

The stop masses are intimately related to the value of the Higgs mass in the MSSM~\cite{Okada:1990vk}--\cite{Degrassi:2002fi}. For a Higgs mass of approximately 125~GeV and equal soft breaking parameters $m_{Q_3} \simeq m_{u_3}$, both stops need to be somewhat heavy, with masses above about 400~GeV~\cite{Carena:2011aa, Arbey:2011ab, Heinemeyer:2011aa, Draper:2011aa}.  In such a case, their loop-effects on the Higgs gluon and photon effective couplings are small, leading to modifications of  at most $\sim 10-20$~\% of the corresponding Higgs production cross section (see for example Ref.~\cite{Benchmark4}).  The Higgs mass constraint, however, can also be satisfied for lighter stops, provided there is a hierarchical relation between the left- and right-handed stop supersymmetry breaking mass parameters, $m_{Q_3} \gg m_{u_3}$~\footnote{   We shall always assume that the left-handed stop mass parameter is larger than the right-handed one. This is because in the small $m_{Q_3}$ case, light sbottoms will appear in the spectrum, which tend to be in conflict with current LHC searches. }. In both cases, a large stop mixing parameter, $A_t$, and a moderate to large value of $\tan\beta$ are required. 

The stop masses, for $X_t \simeq m_{Q_3}$ and $m_{Q_3}\gg m_{u_3}$, are approximately given by
\begin{eqnarray}
m_{\tilde{t}_1}^2 &\simeq& m_{u_3}^2 + m_t^2 \left( 1 - \frac{X_t^2}{m_{Q_3}^2} \right)\;,\\
m_{\tilde{t}_2}^2 &\simeq& m_{Q_3}^2 + m_t^2 \left( 1 + \frac{X_t^2}{m_{Q_3}^2} \right)\;,
\end{eqnarray}
where $X_t =( A_t -\mu/\tan \beta)$. The value of the light stop mass is then given approximately by $m_{u_3}$ and the value of heavier stop mass by $m_{Q_3}$. For stop masses larger than the Higgs mass, their loop contributions to the $\gamma\gamma$ or $gg$ amplitude are approximately proportional to~\cite{Stopsglueglue} 
\begin{equation}
\delta A^{\tilde{t}}_{\gamma \gamma,gg} \propto \frac{m_t^2}{m_{\tilde{t}_1}^2 m_{\tilde{t}_2}^2} \left( m_{\tilde{t}_1}^2 + m_{\tilde{t}_2}^2 - X_t^2\right).
\label{loopamp}
\end{equation}
Hence, for values of the mixing parameter $X_t^2>(<)(m_{\tilde{t}_1}^2 + m_{\tilde{t}_2}^2)$, the stops lead to a reduction (enhancement) of the gluon-gluon Higgs production and an enhancement (reduction) of the Higgs to diphoton decay width. In particular, in the presence of a large hierarchy for the soft masses, $m_{Q_3} \gg m_{u_3}$,  and for large values of $\tan\beta$, the stop loop effects depend dominantly on the relative magnitude of  $A_t$ with respect to  $m_{Q_3}$.  Eq.~(\ref{loopamp}) provides a good parametrization of the stop effects, but it underestimates the stop contributions when their masses are of the order of or smaller than the Higgs mass. Specifically for stop masses of the order of a 100~GeV, they are approximately 30\% larger than the value suggested by Eq.~(\ref{loopamp}).

For reference, we note that we use a normalization in which the  SM contributions to these amplitudes are $\delta A^{t}_{gg} \simeq 4$ and  $\delta A^{W,t}_{\gamma\gamma} \simeq -13 $.
 The stop contributions to the gluon fusion amplitude are approximately given by  $\delta A^{\tilde{t}}_{\gamma \gamma,gg}$, Eq.~(\ref{loopamp}), while those to the $\gamma\gamma$ amplitude are approximately given by $(8/9 \; \delta A^{\tilde{t}}_{\gamma \gamma,gg})$.
 
Two comments are in order:
\begin{itemize}
\item If the stop contribution is of the same sign as the top contribution,  it adds to the gluon fusion amplitude; however it will then contribute to the suppression of the dominant $W$ amplitude in $\gamma\gamma$, and vice versa; 
\item Comparing the relative magnitudes of the SM and stop contributions, we note that the stop effects on the gluon fusion amplitude are approximately a factor of 3.5 larger than their effects on the $\gamma\gamma$ amplitude, normalized to their SM values.   
\end{itemize}

\begin{figure}
\begin{center}
\begin{tabular}{c c}
(i) & (ii) \\
\includegraphics[width=0.48\textwidth]{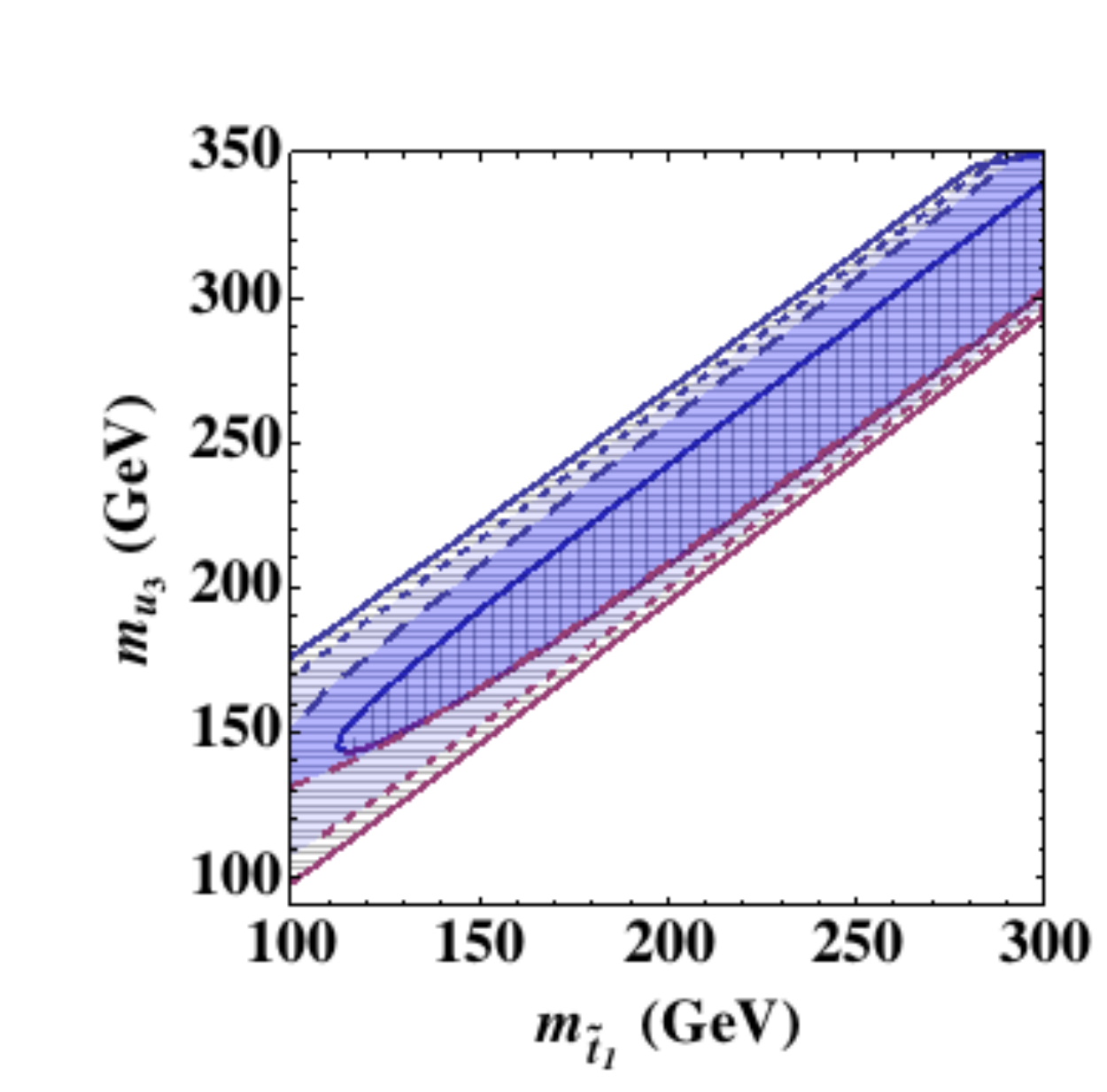}  &
\includegraphics[width=0.48\textwidth]{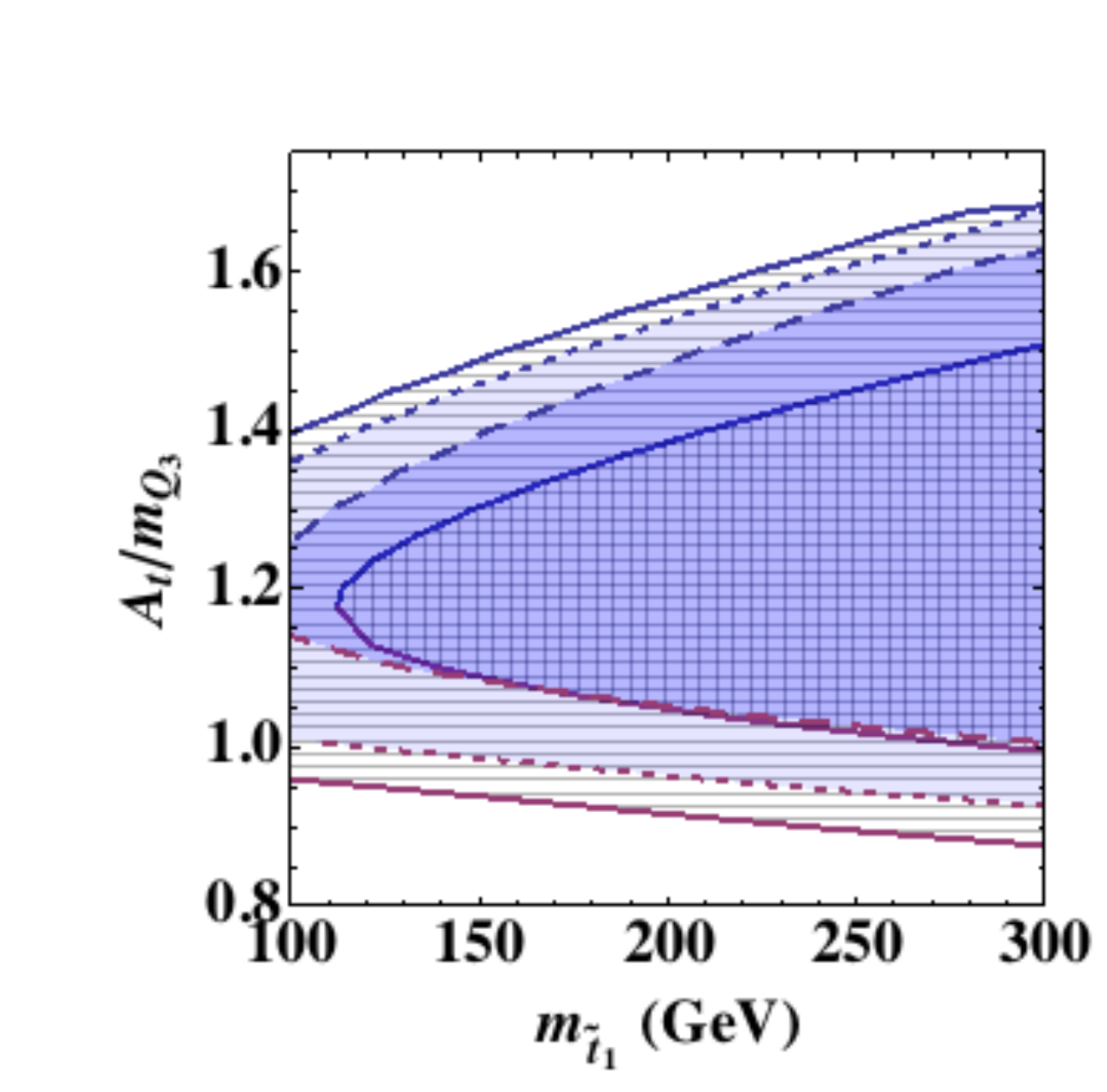} 
\end{tabular}
\end{center}
\caption{Stop mass parameters as a function of the lightest stop mass for the four scenarios listed in Tab.~\ref{table:cases}. Cases (a) and (b) are shown in the two shaded regions bounded by dashed and dotted lines respectively. Cases (c) and (d) are represented by horizontal and vertical hatching respectively. For each value of $m_{u_3}$, values of $A_t$ are such that the computed Higgs mass is in the range 122.5~GeV~$ < m_h < 128.5$~GeV. This range represents a 3 GeV theoretical uncertainty in the $m_h$  computation. The blue contours denote larger values of  $A_t$  and the red contours correspond to the lower values of $A_t$ for a fixed Higgs mass.  }
\label{fig:stoppar}
\end{figure}

\begin{figure}
\begin{center}
\begin{tabular}{c c}
(i) & (ii)\\
\includegraphics[width=0.48\textwidth]{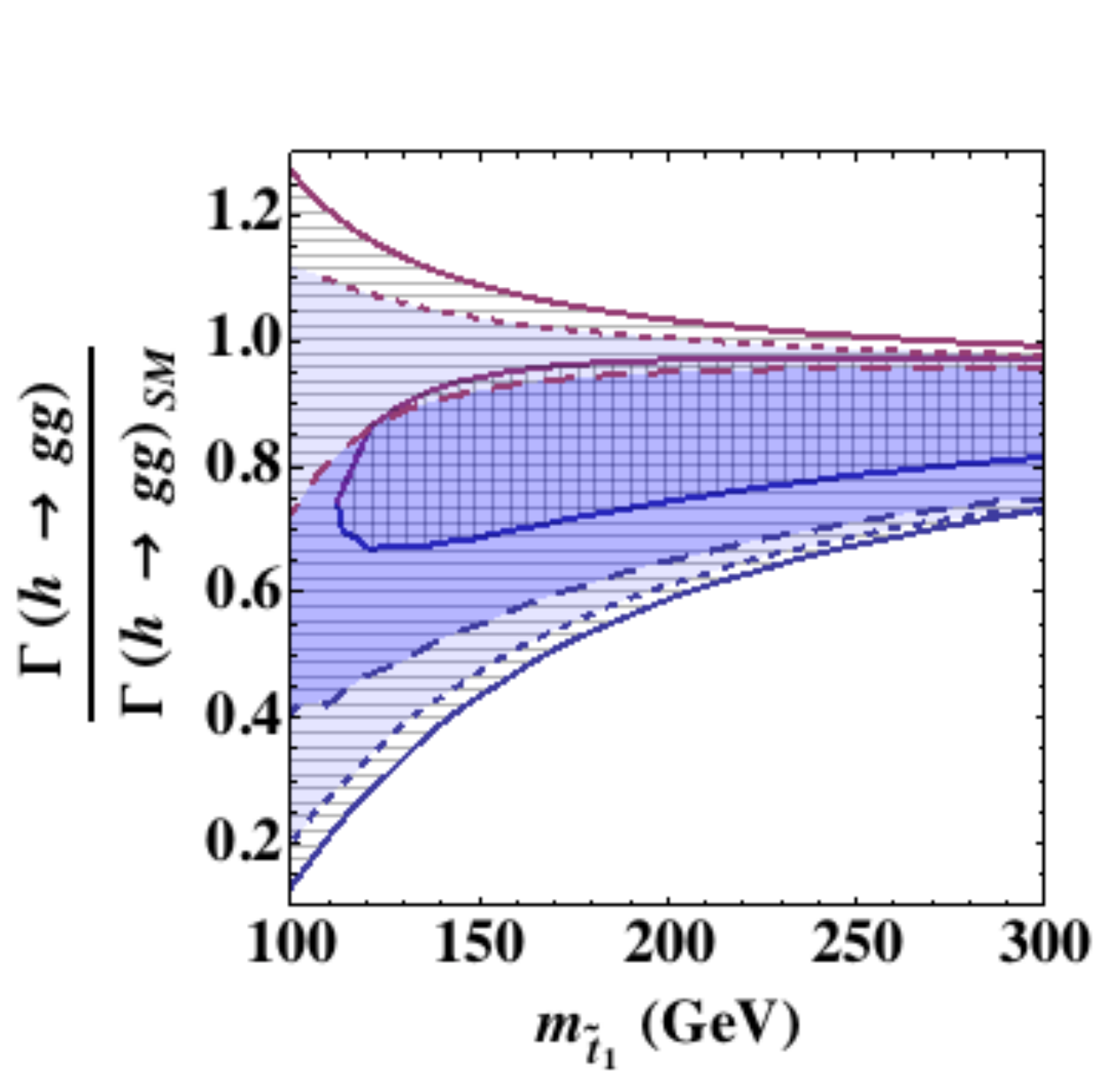}  &
\includegraphics[width=0.48\textwidth]{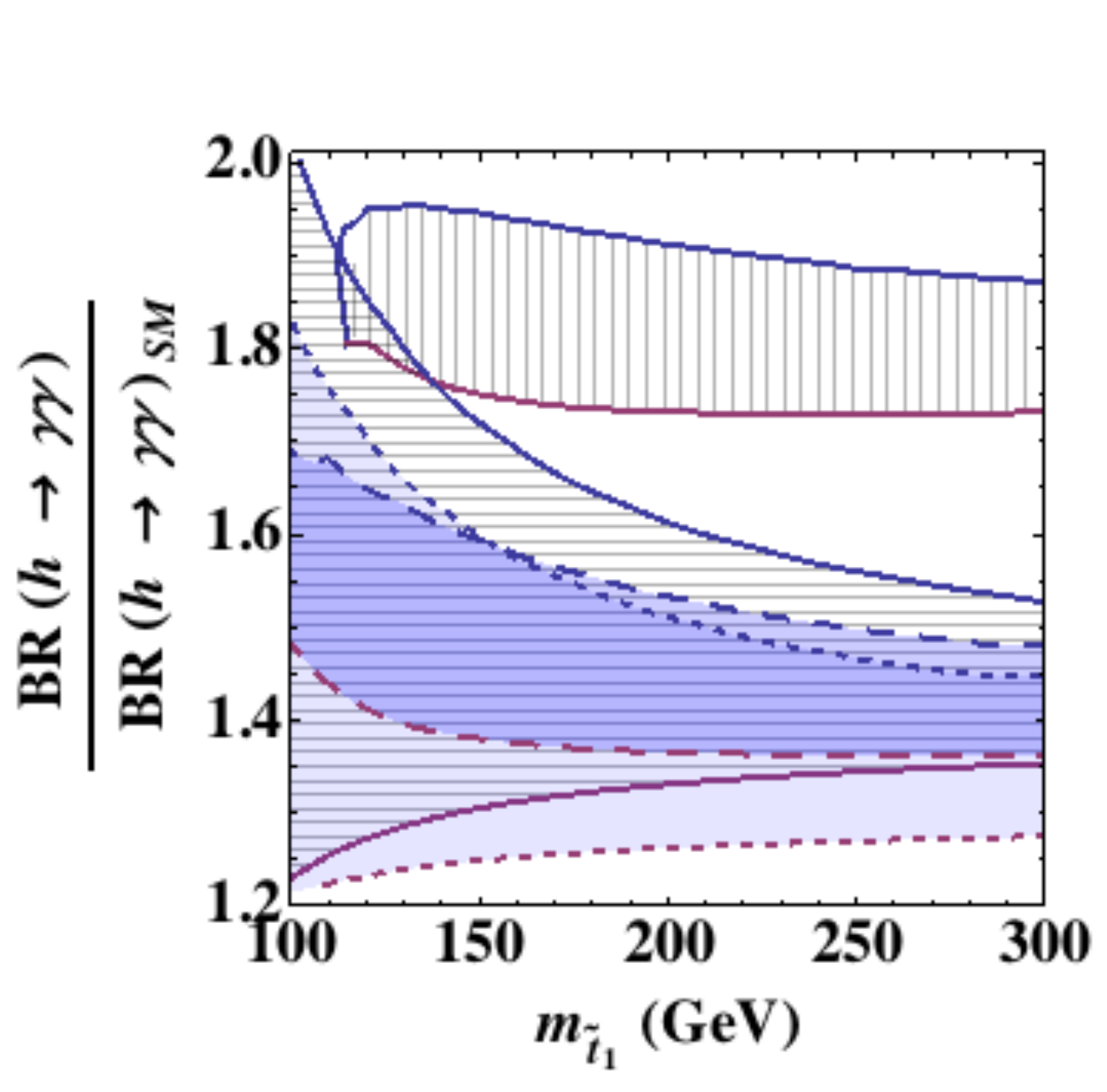}  
\end{tabular}
(iii) \\
\includegraphics[width=0.48\textwidth]{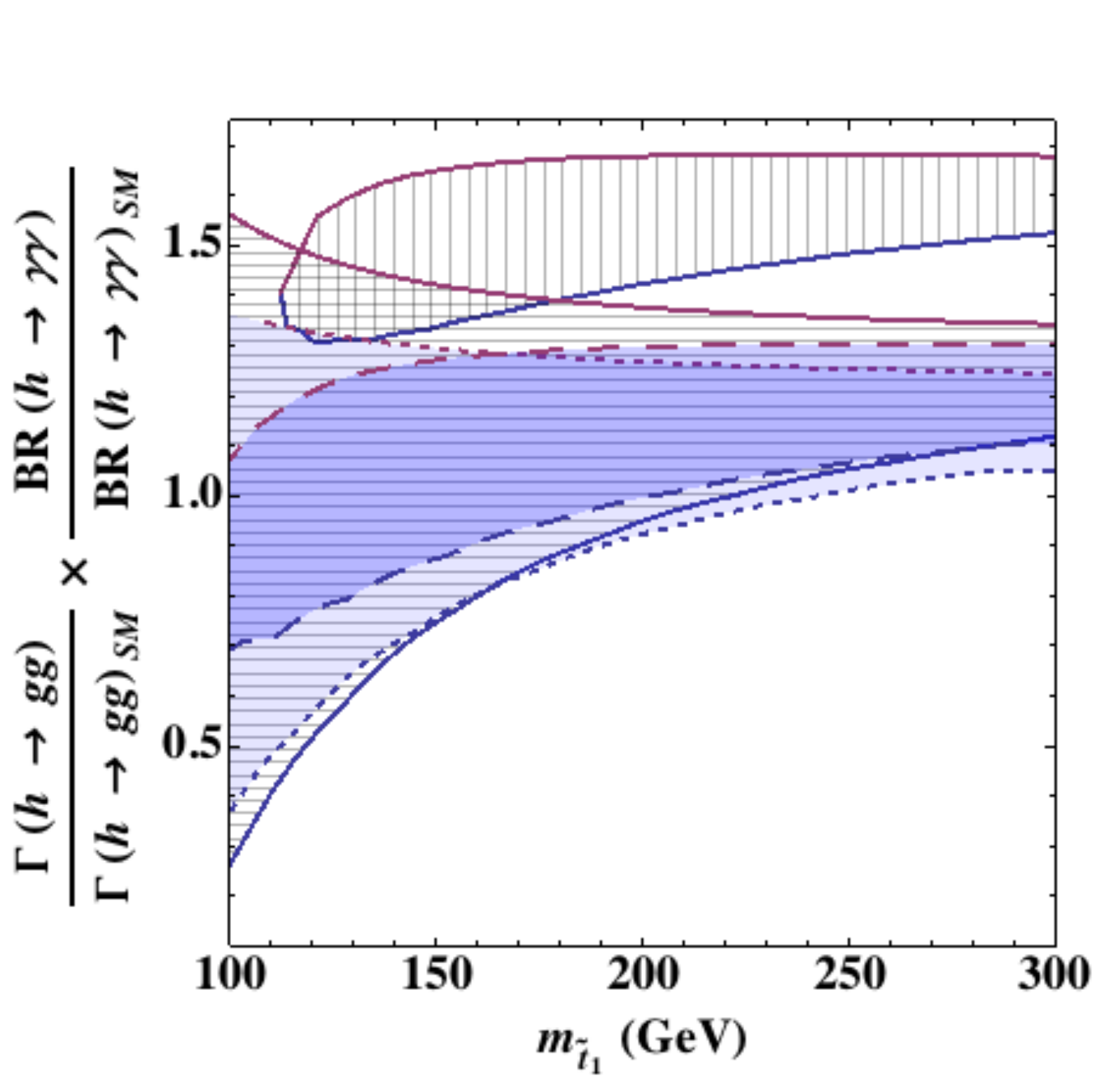}
\end{center}
\caption{ (i) Higgs production via gluon fusion and (ii) its branching ratio into $\gamma \gamma$, normalized to the SM value, as a function of the lightest stop mass for the cases listed in Tab.~\ref{table:cases}. (iii) The $\gamma \gamma$ rate, again normalized to the SM value,  as a function of the lightest stop mass. We use the same conventions as described in Fig.~\ref{fig:stoppar}.}
\label{fig:ggandgamma}
\end{figure}

\begin{table}[h!]
\caption{Parameters defining the different scenarios shown in figures. }
\centering
\begin{tabular}{| l | c |  c | c | c | c | c | c | c |}
\hline\hline
 \;\;\;\;\;\;Cases & $\;\; \tan \beta\;$ & $\;\;m_{\tilde{\tau}_1}$ (GeV) & $\;\;m_{e_3}$ (GeV) & $\;\;\mu$ (GeV) &$\;\; m_{Q_3}$ (TeV) & $\;\;A_\tau$ (TeV) & $\;\;m_{A}$ (TeV)\\
 \hline
 (a) Shaded dashed & 70 & 95 & 250 & 380 & 2 & 0 & 2 \\
 \hline
 (b) Shaded dotted & 70 & 95 & 230 & 320 & 2  & 1 & 1 \\
 \hline
 (c) Horizontal hatch & 105 & 95 & 240 & 225 & 2 & 1 & 1\\
 \hline
 (d) Vertical hatch & 70 & 100 & 300 & 575 & 3 & 1.5 & 1\\
 \hline
\end{tabular}
\label{table:cases}
\end{table}

We present our numerical results using four example scenarios listed in Tab.~\ref{table:cases}, which were analyzed using \texttt{CPsuperH}~\cite{CPsuperH}. Fig.~\ref{fig:stoppar} shows the stop mass parameters  as a function of the lightest stop mass: (i) $m_{u_3}$  and  (ii) the value of $A_t/m_{Q_3}$ needed to obtain the lightest Higgs mass, $m_h \simeq 125.5$~GeV, within a 3~GeV theoretical uncertainty. Note that, for simplicity we  will always consider positive values of $A_t$. Fig.~\ref{fig:ggandgamma} shows the corresponding gluon fusion rate, the Higgs diphoton decay branching ratio and the Higgs-induced diphoton production rate in the gluon fusion channel as a function of the lightest stop mass. The staus are always kept light, and highly mixed, so even for large values of the stop masses, the BR$(h\to \gamma\gamma)$ remains enhanced. 

For a given $m_{Q_3} \gg m_{u_3}$, there are two solutions of positive $A_t$ for each Higgs mass. As discussed earlier, the gluon fusion rate depends on the ratio of $A_t$ to $m_{Q_3}$ and  larger values of $A_t$ lead to smaller  Higgs production rates via gluon fusion. The two solutions are shown with two different colored boundaries in Figs.~\ref{fig:stoppar} and~\ref{fig:ggandgamma}: larger values of $A_t$ are denoted by blue borders and smaller values of $A_t$ by red  borders (the boundaries represent $m_h = 122.5$~GeV).   Since the staus give a negative contribution, proportional to $\left(\mu \tan\beta/m_{e_3}\right)^4$,  to the lightest CP-even Higgs mass (see, for instance, Ref.~\cite{Carena:2012gp})~\footnote{Strictly speaking the corrections depend on $\tan\beta_{\rm eff} = \tan\beta/(1 + \Delta_\tau)$, where $\Delta_\tau$ is the $\tau$ mass threshold correction to be discussed in Sec.~IV.}, the stop mass parameters necessary to obtain consistency with the observed Higgs mass will depend on the stau mass parameters. In particular, for a given value of $m_{Q_3}$, increasing the value of $(\mu\tan\beta/m_{e_3})$ implies that the two solutions for a consistent Higgs mass are obtained for increasingly larger and smaller values of $(A_t/m_{Q_3})$, respectively. This leads to interesting effects in the Higgs gluon fusion production rate.

In scenario (a) (dark blue shaded region in Figs.~\ref{fig:stoppar} and~\ref{fig:ggandgamma}), $m_A=2$~TeV and hence we are effectively in the decoupling regime, in which the Higgs mixing effects are very small.  As can be seen from Fig.~\ref{fig:ggandgamma}, large variations of the gluon fusion production rate may be obtained for stop masses below 200~GeV, with the largest variations corresponding to the larger solution for $A_t$ (blue border).  This can be understood by looking at the corresponding ratios of $(A_t/m_{Q_3})$ shown in Fig.~\ref{fig:stoppar} (ii). The ratio furthest away from 1 leads to the strongest effects. This is in complete agreement with our previous discussion of the stop effects on gluon fusion depending on the ratio of $(A_t/m_{Q_3})$.  The correlated variations of the branching ratio of the Higgs diphoton decay are also relevant in this stop mass regime.  Note that in this case the values of $(A_t/m_{Q_3})$ are such that gluon fusion is always suppressed leading to at most a 30\% enhancement in the Higgs-induced diphoton production rate in gluon fusion processes.

In the second scenario, listed as (b) in Tab.~\ref{table:cases} (light blue shaded region in Figs.~\ref{fig:stoppar} and~\ref{fig:ggandgamma}), $m_A=A_\tau=1$~TeV, and therefore  there can be relevant effects from Higgs mixing induced by stau box-loops. Unlike in scenario (a), due to the smaller negative stau contributions to $m_h$,  here we see that both values of $A_t < (>)~ m_{Q_3}$ are possible, leading to an enhancement~(suppression) of  the gluon fusion Higgs production rate. However, as for (a), the strongest effects are for the larger value of $A_t$. The correlated variations of the Higgs diphoton branching ratio are again also significant.  For the lightest values of the stop masses, we see enhancements of the Higgs-induced diphoton rate in gluon fusion production of up to about 40\% and suppressions of over 50\% compared to the SM expectations.

Scenario (c) (horizontally hatched region in Figs.~\ref{fig:stoppar} and~\ref{fig:ggandgamma}) breaks the requirement of perturbativity up to the GUT scale: $\tan \beta=105$ and $m_A = A_\tau =1$~TeV.  The larger value of $\tan \beta$ allows for a larger $(\mu \tan\beta)$ consistent with vacuum stability~\cite{Stability},  as well as larger Higgs mixing effects (see Sec.~IV), leading to a further enhancement of the Higgs diphoton decay branching ratio of about 20\% compared to scenario (b). The gluon fusion production rate remains very similar to the one in scenario (b), although larger variations are observed consistent with the larger range of $A_t$ values necessary to compensate for the negative stau contributions to $m_h$. 

In all three of the cases above (a, b and c), $\mu$ is always chosen such that the vacuum remains stable, which constrains the largest possible enhancement to the Higgs diphoton production rate to be about 50\%. Further, if we demand perturbativity up to the GUT scale, Higgs mixing effects on the Higgs decay to diphotons  are small.  This is because, for positive values of $(\mu A_\tau)$, $|\mu\tan \beta|$ is forced to be small to stabilize the vacuum~\footnote{We will always assume that $\mu$ is positive.}. However, if we assume that the vacuum may be stabilized by some unknown mechanism for larger values of $\mu$ and $A_\tau$, we can obtain a larger diphoton enhancement for  smaller values of $\tan \beta$ consistent with perturbativity bounds.  Such a scenario,  which is presented for completeness, is analyzed in case (d) (vertically hatched  region in Figs.~\ref{fig:stoppar} and~\ref{fig:ggandgamma}).  We chose a larger value of $m_{Q_3}$ compared to the other scenarios, in order to compensate for the larger negative stau contribution to the lightest CP-even Higgs mass. Even then, the required values of the Higgs mass may not be obtained for stop masses lighter than about 110 GeV.  For this set of parameters the gluon fusion production rate is always smaller than in the SM, with at most a 35\% suppression of this quantity. However, since now  $A_\tau > m_A$ together with a larger $\mu$ compared to the other cases, we have a sizable contribution to the $\gamma\gamma$ rate from Higgs mixing effects as well as from  light stau loops. Therefore, as expected one sees a very large enhancement in the diphoton branching ratio leading to a sizable increase in the Higgs-induced diphoton rate in gluon fusion production, ranging from 30 to 70\%.

 Note that, for the scenarios discussed above, due to the small variation of the Higgs coupling to vector gauge bosons, the Higgs vector boson fusion production rate will be approximately the same as in the SM. Therefore, the ratio of the Higgs-induced diphoton production rate to the SM one in this channel will be given by the corresponding ratio of the Higgs diphoton decay branching ratio.  
Figs.~\ref{fig:ggandgamma} (ii) and (iii) show striking differences between the diphoton branching ratio and the diphoton production rate via gluon fusion, normalized to their SM values. Therefore,
all of these scenarios highlight that if in the future a discrepancy is measured between the diphoton rate from gluon fusion vs. vector boson fusion, it could be evidence for the existence of a very light stop.

\subsection{Light Stop Phenomenology}
\label{stoplimits}

The presence of light stops, with masses below 250~GeV, is highly restricted by present experimental data. However, the region of masses around the threshold of decay of stops into a top and a neutralino is difficult to explore experimentally and is still allowed (see, however, Ref.~\cite{Drees:2012dd,Alves:2012ft,Han:2012fw,Kilic:2012kw}). For neutralino masses around $m_{\tilde{\chi}_1^0} \simeq 30$--$80$~GeV~\cite{Carena:2012gp,NausheenKaty}, consistent with the obtention of the proper Dark Matter density in the light stau scenario, the threshold mass is about 200--250~GeV. Below this threshold, if the chargino is light enough, $m_{\chi^+} < (m_{\tilde{t}_1} - m_b)$, the decay channel $(\tilde{t}_1 \to \tilde{\chi}^+ b)$ could also be relevant. However,  in this section we only consider $m_{\tilde{\chi}^+} > m_{\tilde{t}_1}$.  Therefore, if the stop is lighter, $(m_W + m_b + m_{\tilde{\chi}_1^0} )<m_{\tilde{t}_1} < (m_t + m_{\tilde{\chi}_1^0})$,  it can decay through the 3-body decay channel, $(\tilde{t}_1~\to~b W \tilde{\chi}^0_1)$, mediated by an off-shell chargino or top quark. Finally, for very light stops, $m_{\tilde{t}_1}  <( m_W + m_b + m_{\tilde{\chi}_1^0}) $, the 2-body loop induced decay,  $(\tilde{t}_1 \to  c \tilde{\chi}^0_1)$, typically dominates~\footnote{The 4-body decay mode, $(\tilde{t}_1  \to \tilde{\chi}_1^0 b \ell \nu (\text{or } q \bar q'))$, may be important for $m_{\tilde{t}_1} < 80$ GeV in our scenario~(see however Ref.~\cite{Delgado:2012eu}). We will not consider this case in our paper, since  such small stop masses are generically ruled out by LEP. }. 

\begin{figure}[t]
\begin{center}
\begin{tabular}{c c}
\includegraphics[width=0.48\textwidth]{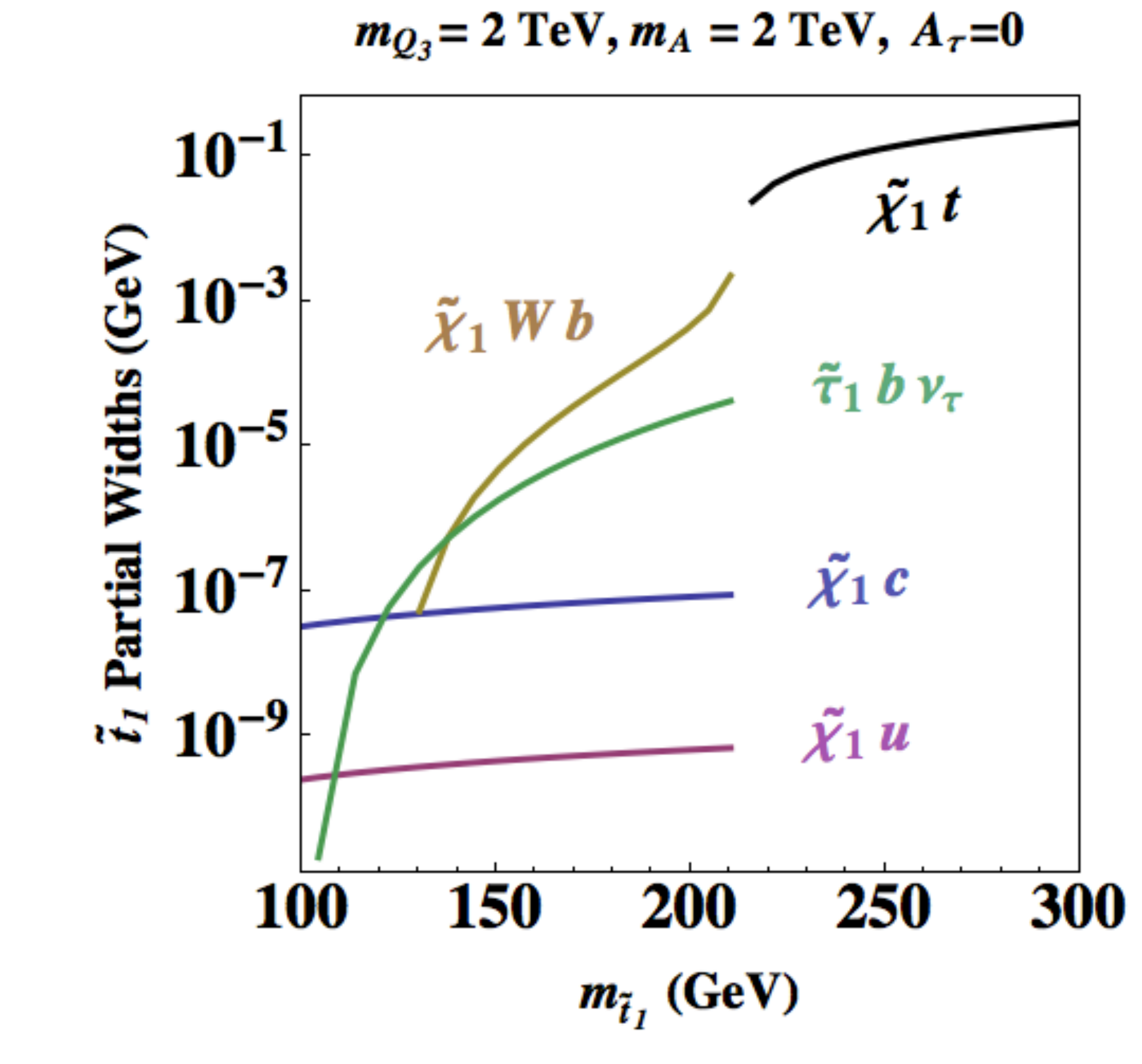}  &
\includegraphics[width=0.48\textwidth]{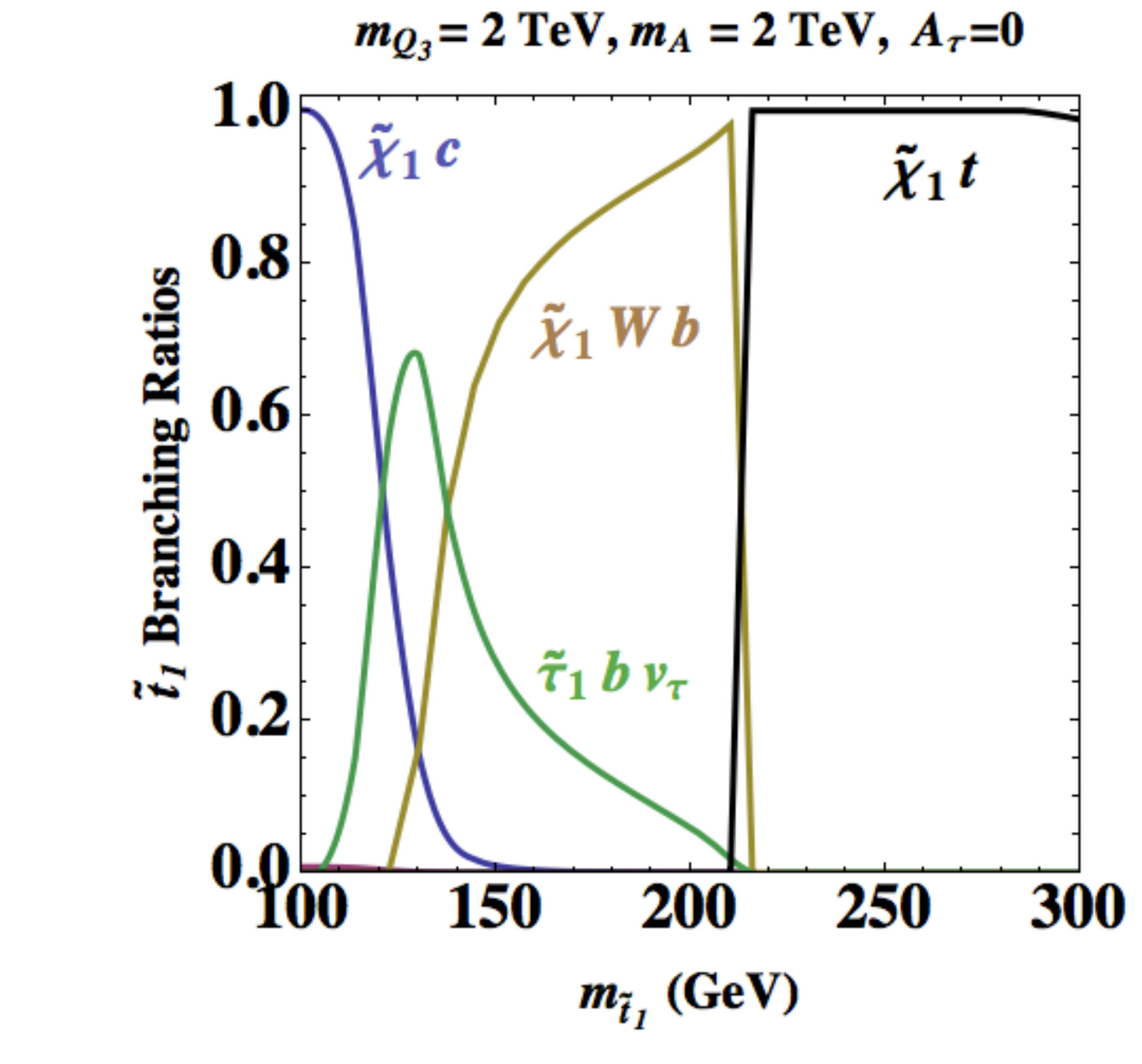}  
\end{tabular}
\end{center}
\caption{Stop partial widths and branching ratios for $\tan \beta =70$, $m_{Q_3} = 2$~TeV, $m_A=2$ TeV, $A_\tau=A_b=0$, $m_{L_3}=m_{e_3}=260$ GeV and $\mu=387$ GeV  corresponding to lightest $m_{\tilde{\tau}_1} \sim 95$~GeV. The gaugino masses will always be set to $M_1=40$~GeV, $M_2=300$~GeV and $M_3=1200$~GeV. For each stop mass corresponding to a given value of $m_{u_3}$, $A_t$ is such that the Higgs mass is $m_h\sim 125$~GeV. }
\label{stopdecays}
\end{figure}

The presence of light staus can open up a new decay channel, ($\tilde{t}_1 \to \tilde{\tau}_1^+ \nu_{\tau} b)$. This new channel could be the dominant one whenever the 3-body decay channel,  $(\tilde{t}_1 \to \tilde{\chi}_1^0 W^+ b)$, is kinematically suppressed.  In our analysis, we  choose a stau mass of 95~GeV and a neutralino mass of 40~GeV. This would be approximately consistent with what would be required to obtain the proper relic density if the neutralino annihilation rate was dominated by the s-channel $Z$ interchange contribution~\cite{Carena:2012gp,NausheenKaty}. For these values of the stau and neutralino masses, the 3-body decay channel gets suppressed for stop masses in the range 110--140~GeV. In Fig.~\ref{stopdecays}  we present the stop branching ratios showing the  appearance of this new channel in the presence of light staus. The slepton parameters are chosen {as in scenario (a)}. The left-handed stop mass parameter, $m_{Q_3}$, is fixed to 2 TeV and the values of $m_{u_3}$ and $A_t$  are always chosen such that the Higgs mass is $m_h \sim 125$~GeV. The value of the trilinear term, $A_b$, plays a relevant role in the stop decay rate into a charm and a neutralino~\cite{Hikasa:1987db}. Assuming a radiative generation of the stop coupling to charm and neutralino in minimal flavor violating models, negative values of $A_b$, with $-m_{Q_3}< A_b < 0$, tend to enhance the rate with respect to the $A_b =0$ case, while sizable and positive values of $A_b$ tend to suppress it. We choose $A_b=0$ here as an illustrative case. 

In the rest of this section, we recast existing stop searches into limits for the light stops we analyze in this paper. The limits presented here are based on parton level simulations done with {\texttt Madgraph5}~\cite{Alwall:2011uj}. We stress that our analysis is simplistic, and can
by no means replace a full collider study. Nevertheless, it is interesting to highlight the prospects of 
probing these light stops at the LHC, since they are able to significantly alter the Higgs phenomenology while being consistent with the measured Higgs mass.

\begin{enumerate}
\item Let us first consider a scenario with a very light stop, $m_{\tilde t_1}\lesssim 120$ GeV. The main decay mode is given by $(\tilde t_1\to\tilde\chi_1 c)$. The most stringent bound on this channel comes from a CDF analysis~\cite{Aaltonen:2012tq} based on 2.6 fb$^{-1}$ of data. For neutralino masses of around 40~GeV, assuming  BR$(\tilde{t}_1 \to \tilde{\chi}^0_1 c) =100 \%$, the range $m_{\tilde{t}_1} \lesssim 124$ GeV is excluded. This bound is slightly weakened in our model, since in a narrow mass range around $m_{\tilde{t}_1} \simeq 120$ GeV, the decay mode $(\tilde t\to\tilde\tau\nu_\tau b)$ is competing with $(\tilde t\to \tilde{\chi}^0_1\, c)$. However stops below $\simeq (115-120)$ GeV are still excluded by the CDF $(\tilde{\chi}^0_1 \, c)$ search. Additionally, it has been shown~\cite{Krizka:2012ah} that ATLAS and CMS monojet, jets+MET, Razor, and $M_{T2}$ analyses, based on $\sim 5\,\rm{fb}^{-1}$ 7 TeV data, give constraints on the parameter space for $m_{\tilde t_1}\lesssim 120$ GeV that are comparable to those coming from the Tevatron  $(\tilde\chi^0_1 c)$ search.

\item Next, we consider the mass range $120 $ GeV $ \lesssim m_{\tilde t_1} \lesssim 160 $ GeV. In this interval the decay  $(\tilde{t} \to \tilde\tau^+_1\nu_\tau b)$ is kinematically open. There is no specific stop experimental search in this decay mode. In Ref.~\cite{Yu:2012kj} it has been shown that, if the stau and the LSP are nearly degenerate in mass, the $\tau$ lepton from the stau decay 
would be soft and difficult to identify. In this case, important bounds may come from ($b$-jets+MET) searches.

In our scenario, the splitting between the LSP and the stau is sizable. Hence these bounds should not be applied. However, the final state, ($\tau b\ + $ MET), is the same as {\ns for} the top decay mode $(t \to b W \to b \tau \nu_\tau)$. Moreover, due to the fact that in this region of parameter space the stop mass is close to the top mass, the kinematics would be similar as well. Therefore, the measurement of the $t \bar t$ production cross section can set limits on this channel. 
\begin{itemize}
\item We first analyze the case in which both stops decay into $(\tilde{\tau}^+_1b\nu_\tau)$, each leading to a ($\tau b$ + MET) final state. The measurement of the SM $t \bar t$ production cross section in the  ($\tau$ + jets) mode \cite{Abazov:2010pa,cms11004,Collaboration:2012hz,Aad:2012vna,:2012cj} ($pp\to t\bar t$ with $t\to Wb$ and one $W$ decaying leptonically and the other hadronically), does not give important constraints on our scenario since these analyses typically require too many (4-5) final state hard jets. Additionally, there are measurements in the $\ell\ell +$ (at least) two jets mode~\cite{:2012bta,ATLAS:2012aa,cmsllttbar}. These searches, would affect our scenario if both $\tau$'s decay leptonically. However, we have checked that so far these searches do not give an important limit on our scenario, since the stop signal rate in this channel is very small. On the other hand, there are dedicated searches where the $W$s from the $t\bar{t}$ decay produce one tagged hadronic tau and one additional lepton~\cite{:2012xh,Chatrchyan:2012vs}. These searches give us much stronger limits. 

We performed a simple parton level recast of the latter ATLAS and CMS $t\bar t$ cross section measurements. In the (120-160) GeV mass range, the number of events we obtain is of the order of the 1-$\sigma$ error on the expected background. Therefore, since the measurement is in good agreement with the SM prediction,  if BR$(\tilde{t}_1 \to \tilde \tau^+_1 \nu_\tau b ) = 100 \%$, the current experimental limit would already be probing a light stop in almost the entire mass range (120-160) GeV at the 1 or 2-$\sigma$ level. However, from Fig.~\ref{stopdecays}, we see that BR$(\tilde{t}_1 \to \tilde\tau^+_1\nu_\tau b) $ is typically less than $\sim 70 \% $.  Therefore, the $t\bar t$ cross section measurement in the $(\tau+\ell)$ channel is still not probing the stop mass range $(120-160)$ GeV~\footnote{Note however, that  the region of stop masses corresponding to the maximum of the branching ratio BR$(\tilde{t}_1 \to \tilde\tau^+_1\nu_\tau b$), $m_{\tilde{t}_1}\sim 130$ GeV, is very close to being probed in this channel.}.

\item Let us now consider the case in which one stop decays to $(\tilde\tau^+_1b\nu_{\tau})$ and the other one to $(Wb\tilde\chi_1^0)$. This case is particularly relevant for stop masses of about 135~GeV  where BR$(\tilde t\to Wb\tilde \chi^0_1)\sim{\rm{BR}}(\tilde t\to \tilde\tau^+_1 \nu_\tau b)$.  We checked that the $t \bar t$ production cross section measurements in the $(\tau +\ell)$ mode~\cite{:2012xh,Chatrchyan:2012vs} and the ($\tau +$ jets) mode~\cite{cms11004} give similar constraints in this region of stop masses. Comparing this case to when both stops decay to $(\tilde\tau^+_1b\nu_{\tau})$,  we note that this decay
mode has a slightly larger signal to background ratio. Therefore, light stops with masses at around 135 GeV will be first tested by $t \bar{t}$ production cross section measurements in the $(\tau +\ell)$ and $(\tau +$ jets) modes. 
\end{itemize}
  
\item Finally, for stops in the mass range  $150 $ GeV $ \lesssim m_{\tilde t_1} \lesssim 210 $ GeV, the dominant decay mode is  $(\tilde{t}_1\to W b \tilde{\chi}_1^0)$. No experimental search has been performed for this 3-body decay. However, several theoretical papers, in the framework of stops NLSP, re-analyze some of the existing Tevatron and LHC analyses to put bounds on this stop decay~\cite{Kats:2011it,Choudhury:2012kn,Krizka:2012ah}.
 
Additionally, constraints can come from recasting searches for stops decaying through the 2-body decay channel: $(\tilde t \to b\tilde \chi^+\to b W^+ \tilde{\chi}_1^0)$, both from ATLAS \cite{ATLAS:166,ATLAS:167,ATLAS:001} and CMS  \cite{CMS:stopbchargino}, by considering off-shell charginos. However, in general the signal acceptance will be rather low. This is because, in comparison with the scenarios considered by the LHC searches, our case predicts different kinematics for the final state particles. Since  both ATLAS and CMS searches assume stop decays into an on-shell chargino, they mainly  focus on a region of parameter space where $(m_{\tilde{\chi}^+} - m_{\tilde{\chi}_1^0}) < m_W$. The chargino then decays into the LSP and an off-shell $W$ boson.  On the other hand,  in our model the decay $(\tilde{t}_1\to W b\tilde{\chi}^0_1)$ proceeds through a 3-body decay mediated by an off-shell chargino or a top quark. The $W$ boson is on-shell in this region of stop masses. Therefore, the leptons produced from the decay of such an on-shell $W$ are in general more energetic. Additionally the missing energy will be smaller in the case of a 3-body decay.

Recent phenomenological analyses suggest~\cite{Krizka:2012ah,Delgado:2012eu} that the most constraining searches are not from dedicated stop searches, but from using LHC analyses with $b$-jet final states and in particular the CMS $b$-jet, Razor, MT2 analyses. Such searches could place strong limits on this scenario in the entire mass range, unless BR$(\tilde{t} \to W b \tilde{\chi}_1^0)$ is significantly suppressed. Stops with masses larger than $\sim$ 140 GeV are therefore ruled out.

\end{enumerate}

To summarize, due to the new decay mode, $(\tilde{t} \to \tilde\tau^+_1\nu_\tau b)$, light stops could evade the current experimental bounds in a narrow mass window, $120\, {\rm{GeV}} \lesssim m_{\tilde{t}_1}\lesssim 140$~GeV.  At the same time, current SM measurements of the $t \bar t$ production in $\tau$ final states are already very close to directly probing this region of parameter space. A dedicated search could therefore explore this possibly interesting light stop signal.

\section{Bottom and Tau  Higgs Decays}
\label{BtauCoup}

\subsection{Higgs Mixing Effects and the Bottom and Tau Higgs Branching Ratios}

In the supersymmetric limit, the bottom quark and the tau lepton couple only to the down-type Higgs, $H_d$, with couplings $h_{b,\tau}$, respectively.  After supersymmetry breaking, both fermions also couple to the up-type Higgs, $H_u$, via loop-induced couplings, $\Delta h _{b,\tau}$.  Hence, the couplings of these fermions to the 
lightest CP-even Higgs are given by~\cite{Carena:1999bh}
\begin{equation}
g_{hbb,h\tau\tau} = -h_{b,\tau} \sin\alpha + \Delta h_{b,\tau} \cos\alpha,
\end{equation}
where $\alpha$ is the CP-even Higgs mixing angle and (-$\sin\alpha$) and $\cos\alpha$ are the projections on $h$ from the real neutral components of $H_{d}$ and $H_u$, respectively. The $b$ and $\tau$ masses are given by~\cite{deltamb,deltamb1,deltamb2,deltamb3}
\begin{eqnarray}
m_{b,\tau} & = & h_{b,\tau} v_d \left( 1 + \tan\beta\; \frac{\Delta h_{b,\tau}}{h_{b,\tau}}\right)\, ,
\nonumber\\
                   &\equiv& h_{b,\tau} v_d \left(1 + \Delta_{b,\tau}\right).
\end{eqnarray}
Hence,
\begin{equation}
g_{hbb,h\tau\tau} =  -\frac{m_{b,\tau} \sin\alpha}{v \cos\beta (1 + \Delta_{b,\tau})} \left( 1 -  \frac{\Delta_{b,\tau}}{ \tan\beta \tan\alpha}\right).
\end{equation}
Close to the decoupling limit, which is when the CP-odd Higgs  mass is very large, and at large values of $\tan\beta$, $\sin\alpha$ is close to $(-\cos\beta)$  and $\cos\alpha \simeq \sin\beta \simeq 1$. The ratio $(\sin \alpha / \cos \beta)$ is then $(\tan \alpha \tan \beta)$, to a very good approximation, and the couplings can be written as:
\be
g_{hbb, h\tau\tau}\sim  \frac{m_{b,\tau}}{v}\left[1+ \frac{|\sin\alpha/\cos \beta| - 1}{1+\Delta_{b,\tau}}\right]\,. \label{btau}
\ee
Note that when $(\sin\alpha \to -\cos\beta)$, the above expression reproduces the SM values. 
We can also see that the suppression or enhancement of the couplings with respect to the SM will depend on whether  $|\sin \alpha/\cos \beta|$ is greater than  or less than 1.  On the other hand, independent of the value of  $|\sin \alpha/\cos \beta|$, we see that larger deviations from the SM couplings are given by smaller values for $(1+\Delta_{b,\tau})$. This implies that positive~(negative) values of $\Delta_{b,\tau}$ would lead to values closer to~(further away from) the SM. As we have shown in Ref.~\cite{Stability},  positive (negative) values of $\Delta_b$ ($\Delta_\tau$) are obtained in our scenario for positive values of $\mu$ and the gauging masses. Therefore, in this case we expect that $g_{hbb}$ will be closer to the SM value than $g_{h\tau\tau}$ for the same set of parameters.

As regards to the ratio of the couplings, since $\Delta_b \neq \Delta_\tau$, this is no longer given by $(m_b/m_\tau)$, as at tree level, but rather by
\begin{equation}
\frac{g_{hbb}}{g_{h\tau\tau}} = \frac{m_b (1 + \Delta_\tau)\left(1-\Delta_b/(\tan\beta \tan\alpha)\right)}
                                                              {m_\tau (1 + \Delta_b)\left(1-\Delta_\tau/(\tan\beta \tan\alpha)\right)}\, .
\end{equation}
If we assume that the loop effects are small, and that the couplings admit an expansion on $\Delta_b$ and $\Delta_\tau$, the ratio of the couplings, normalized to their SM values, can be approximated by
\bea
\left(\frac{g_{hbb}}{g_{h\tau\tau}}\right)_{\rm SM} &\sim& 
1-\left( \Delta_b-\Delta_\tau\right)\left(1-\left|\frac{\cos \beta}{\sin \alpha}\right| ~\right)\,.\label{btaurat}
\eea
We see that the ratio with respect to the SM will also be governed by the value of $|\sin \alpha/\cos \beta|$. However, comparing Eqs.~(\ref{btau}) and (\ref{btaurat}), we see that when $(\Delta_b - \Delta_\tau) >0$ and $|\sin \alpha/\cos \beta| < (>) 1$, the couplings themselves are suppressed (enhanced) compared to the SM, but the ratio of the couplings, is in fact enhanced (suppressed).

\begin{figure}
\begin{center}
\begin{tabular}{c c}
(i) & (ii)\\
\includegraphics[width=0.48\textwidth]{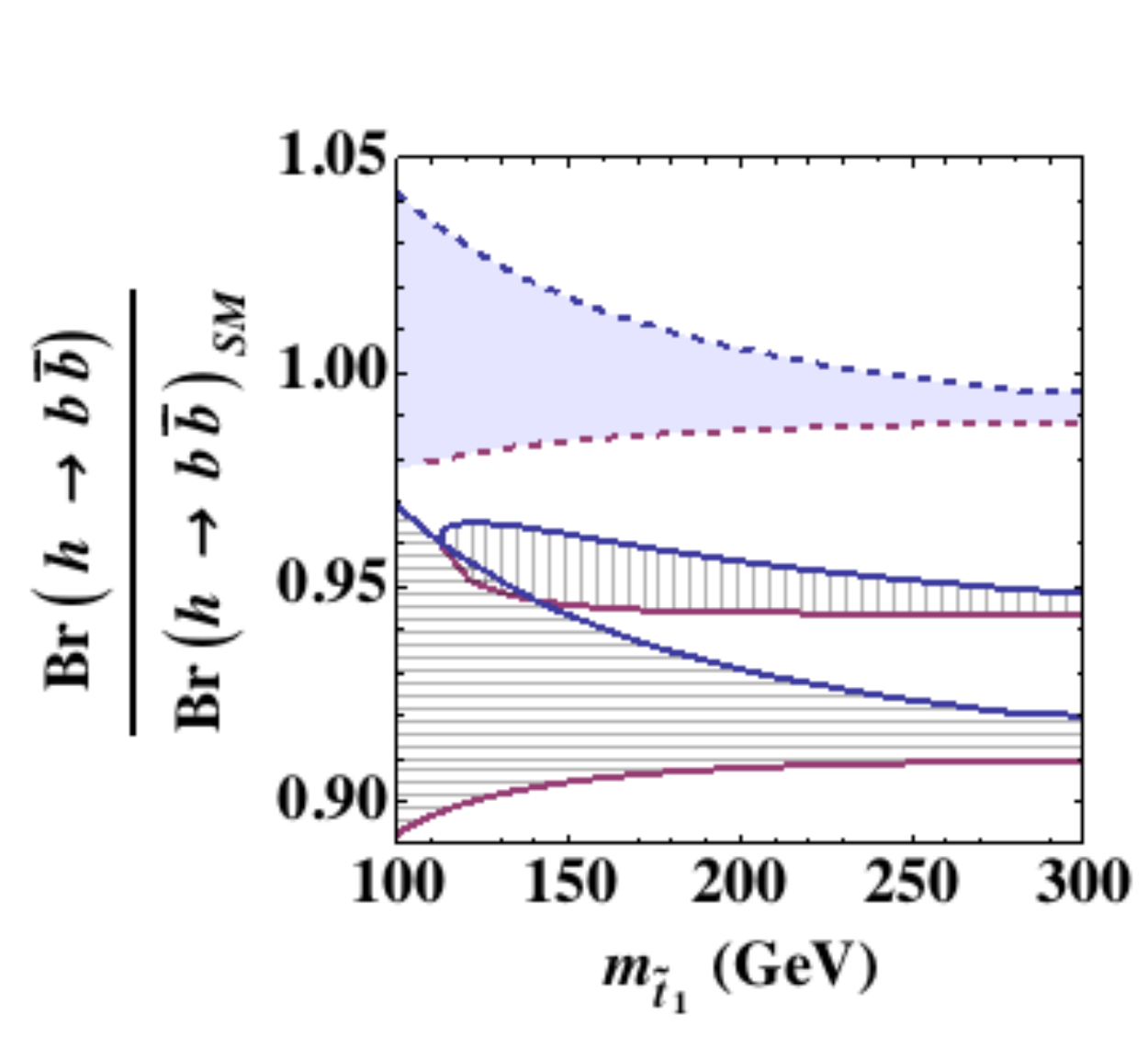}  &
\includegraphics[width=0.48\textwidth]{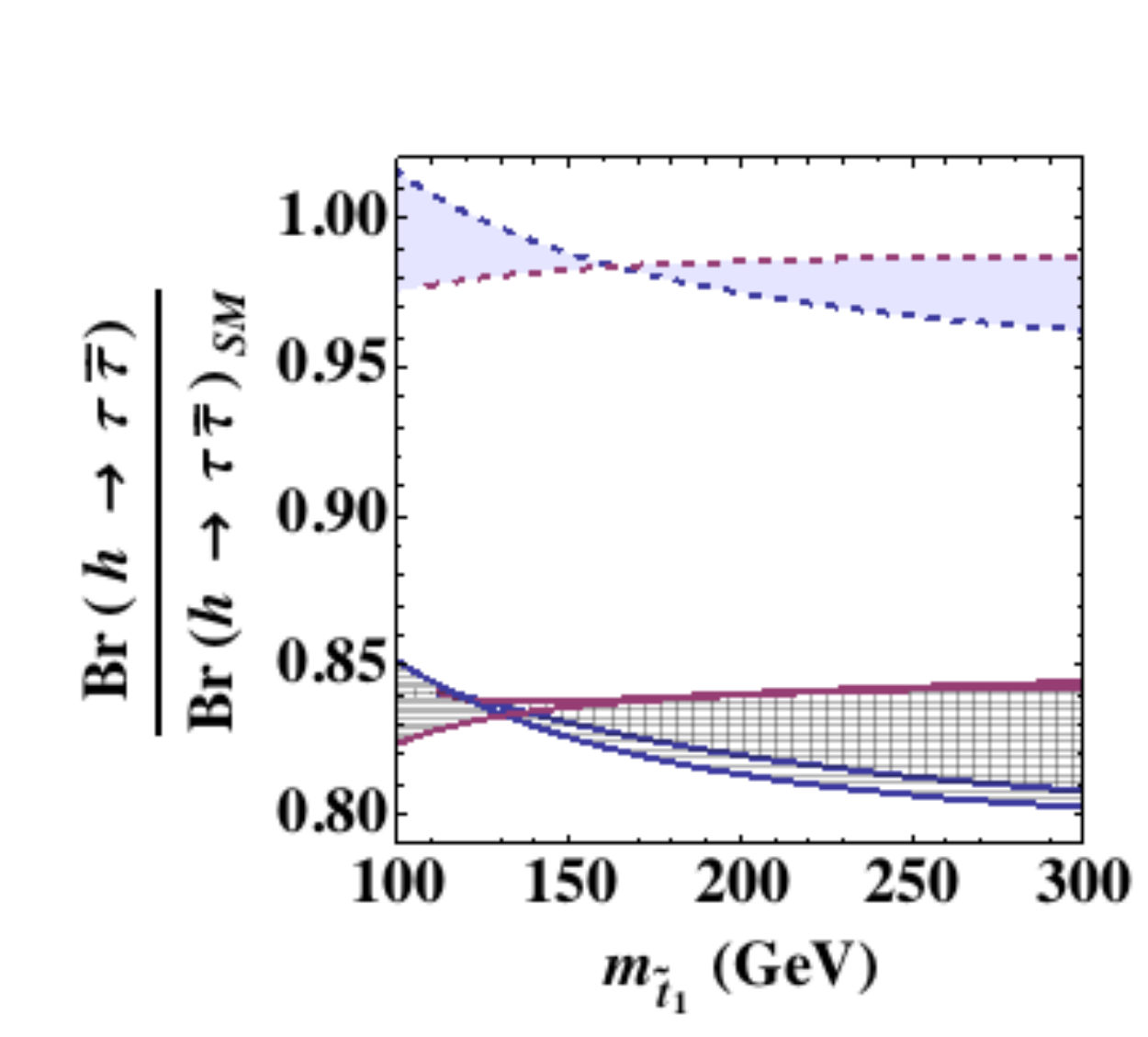}  
\end{tabular}
(iii)\\
\includegraphics[width=0.48\textwidth]{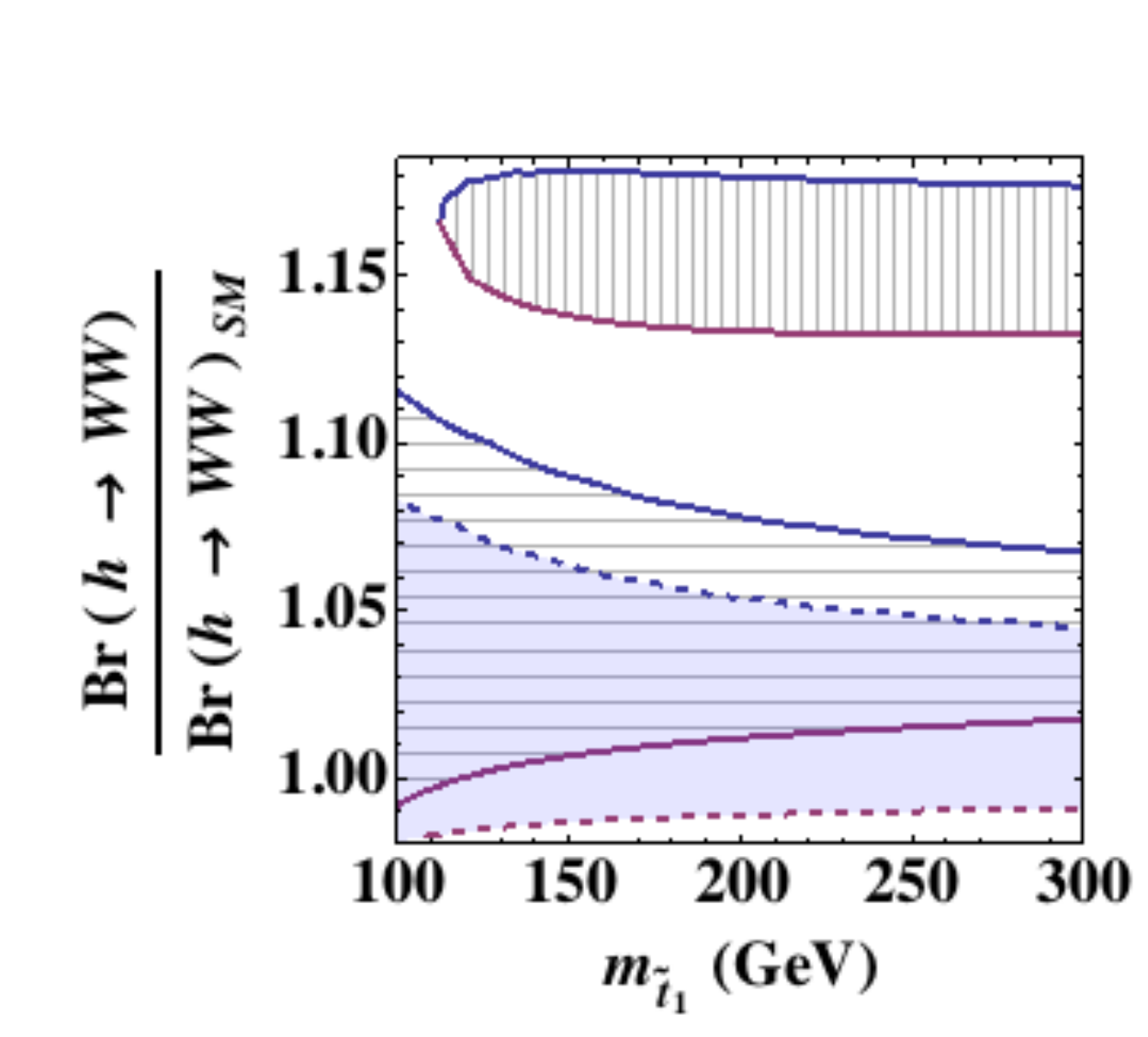}  
\end{center}
\caption{Higgs decay branching ratios into $b\bar b$, $\tau^+\tau^-$ and $W^+W^-$ as a  function of the lightest stop mass for scenarios (b), (c) and (d) presented in Tab.~\ref{table:cases}. The lines, colors and hatching is as described in Fig.~\ref{fig:stoppar} and Tab.~\ref{table:cases}. }.
\label{BrhbbhtautauWW}
\end{figure}

For moderate values of $m_A$ and sizable positive values of $A_\tau$, $|\sin\alpha/\cos \beta|<1$ and therefore there could be a sizable suppression of the coupling $g_{h\tau\tau}$ and a moderate suppression of $g_{hbb}$. If the dominant  decay width, $\Gamma(h \to b\bar{b})$, is suppressed, not only the Higgs to diphoton rate, but also the Higgs vector boson decay branching ratios are enhanced. This in turn implies that, in the presence of light stops, the rate $(gg\to h\to VV)$ can become approximately standard, due to a compensation of the suppression of the gluon fusion rate with the enhancement of the ${\rm{BR}}(h\to VV)$.  

Fig.~\ref{BrhbbhtautauWW} shows the decay branching ratios for $bb$, $\tau^+\tau^-$ and $W^+W^-$ normalized to the SM, for scenarios (b), c) and (d) presented in Tab.~\ref{table:cases}. As expected from the discussion above, the variations are more significant in the $(h \to \tau^+\tau^-)$ case and increase for smaller ratios of $m_A$ to $A_{\tau}$~(compare scenarios (b) and (d)).  While ${\rm{BR}}(h \to b\bar{b})$  remains within 10\% of the SM value, ${\rm{BR}}(h\to \tau^+\tau^-)$ can vary by up to almost 20~\% of  the SM value.  

The variations in   BR$(h\to W^+W^-)$  are mostly induced by changes in the dominant decay width, $\Gamma(h \to b\bar{b})$. However, contrary to expectations, for case (b) one observes an enhancement for both the $WW$ and $b\bar{b}$ branching ratios together. In cases (c) and (d), even when one observes a correlation between the suppression of $b\bar{b}$ and the enhancement of $WW$, we note that  larger suppression in $b\bar{b}$ does not lead to a larger enhancement in $WW$. This is due to the fact that even though the largest partial width is $\Gamma_{b\bar{b}}$, the variation in the partial widths of the Higgs decay into taus and gluons play a relevant role in determining the total width in this region of parameter space. There is a larger variation of the decrease in the partial width for the Higgs decay into gluons and taus for the larger values of $A_t$~(blue border),  with very strong suppressions of the decay into gluons for the smallest stop masses. This leads to a relevant decrease of the total width (and consequently an increase in BR$(h\to W^+W^-)$), beyond the behavior expected from just the variation in the branching ratio into $b\bar{b}$. For the smaller values of $A_t$~(red lines) we have the opposite effect on the total width, since now the partial widths of the Higgs decay into gluons are larger. However, here the enhancements in the Higgs gluon decay width are smaller than the suppressions seen for the larger values of $A_t$. Therefore, there exists an anti-correlation between the total width with $A_t$ in this region of parameters, with  larger value of $A_t$ leading to smaller total widths and vice versa. However, note also that generically, the lower value of $A_t$ corresponds to a smaller variation in the effects on the total width as a function of the stop mass. Therefore, since the partial width of the Higgs decay into $WW$ remains approximately SM like, we see that the $WW$ decay branching ratio exhibits both larger enhancements and larger variations for the larger values of $A_t$.

\subsection{Heavy Higgs Phenomenology}

Since the largest variations in the $\tau\tau$ decay are obtained for the smaller values of $m_A$, a relevant constraint on these NP effects could come from experimental bounds on $m_A$. One has to consider the variation of these bounds in the light stau scenario with respect to the traditional ones coming from $(A,H\to\tau\tau)$ searches. An important effect comes from the fact that for large values of $A_\tau$, the heavy neutral Higgs bosons, $H$ and $A$, have a sizable coupling to $(\tilde\tau_1\tilde\tau_1)$, $(\tilde\tau_2\tilde\tau_2)$ and $(\tilde\tau_1\tilde\tau_2)$ and may therefore decay into final states containing staus. Additionally, relevant effects can occur due to the appearance of  non-negligible  mass correction factors:  a large positive value of $\Delta_b$, and a smaller negative value for $\Delta_\tau$.

At present, the CMS and ATLAS collaborations are setting bounds on the $(m_A-\tan\beta)$ plane up to pseudo-scalar masses of 800 GeV~\cite{LHCtautauMSSM}, using 4.6-4.8 fb$^{-1}$ of data at 8 TeV. In particular for $m_A=800$ GeV the bound on $\tan\beta$ is at around 50. Both experiments present their results for the so called $m_h^{\rm max}$ scenario~\cite{Benchmark4}, with $M_{\rm{SUSY}} = 1$ TeV, $X_t = 2~M_{\rm{SUSY}}$,  $\mu = 200$ GeV,  $M_{\tilde g} = 800$ GeV, $M_2 = 200$ GeV and $M_1$ fixed by the relation $M_1=5/3 M_2 \tan^2\theta$.   Squarks and sleptons are kept at the scale $M_{\rm SUSY}$ and therefore do not contribute to the heavy Higgs decay width. We need to convert the presented experimental limits to the light stau scenario, taking into account the effects outlined above. 

First note that even if the only relevant decays of the heavy Higgs bosons were into $b\bar{b}$ and $\tau^+\tau^-$, the searches for heavy Higgses decaying into $\tau^+ \tau^-$ could have a significant dependence on $\Delta_{b, \tau}$.  The reason for this is that the production cross section of $A$ and $H$ at hadron colliders is induced by the couplings of the heavy Higgs bosons to $b$, and is therefore proportional to
\be
h_b^2    \simeq     \frac{m_b^2 \tan\beta^2}{ v^2 (1 +\Delta_b)^2  }\;.
\ee
On the other hand, the branching ratio of the decay into $\tau^+ \tau^-$ is proportional to
\be
{\rm{BR}}(H,A\to \tau^+ \tau^-) \propto \frac{h_\tau^2}{3 h_b^2 + h_\tau^2}\;,
\ee
where the 3 comes from the number of bottom-quark colors and $h_\tau \simeq m_\tau \tan\beta/[v (1 + \Delta_\tau)]$. In terms of the loop corrections, this can be written as:
\be
\label{brnostau}
{\rm{BR}}(H,A\to \tau^+ \tau^-) \propto  \left[ 3~ \frac{m_b^2}{m_{\tau}^2}\frac{ (1 + \Delta_\tau)^2}{(1 + \Delta_b)^2 } + 1 \right]^{-1}\,.
\ee
The heavy Higgs production cross section times the branching ratio of the Higgs decay into a pair of $\tau$'s then becomes proportional to
\be
\sigma(pp\to H,A) \times {\rm{BR}}(H,A\to \tau^+ \tau^-) \propto \frac{m_b^2 \tan^2\beta}{(3 \frac{m_b^2}{m_\tau^2} (1 + \Delta_\tau)^2 + (1 + \Delta_b)^2)}\;.
\label{prodnostau}
\ee
Therefore for $\Delta_\tau = 0$, $\Delta_b$ appears as a subdominant correction. However, for the specific parameter regions we are looking at, it can still give corrections of about 15-20\%. Further, $\Delta_\tau$ negative increases $h_\tau$ and hence the $\tau^+ \tau^-$ decay partial width. This can be seen by the fact that a negative $\Delta_\tau$ reduces the denominator in Eq.~(\ref{prodnostau}), and hence both $\Delta_{b,\tau}$ start having a  larger impact on the $\tau$ production cross section. 

If one now includes decays into light staus, the heavy Higgs production rate will not change, but, the branching ratio of the decay into $\tau$'s will be reduced. Ignoring phase space factors, the heavy Higgs decay width into staus is proportional to
\be
\sum_{i =1,2} \Gamma(H\to \tilde{\tau}_i\tilde{\tau}_i) \simeq 2 \Gamma(A \to \tilde{\tau}_1 \tilde{\tau}_2) \propto \frac{h_\tau^2 A_\tau^2}{m_A}\;,
\ee
instead of to $(h_\tau^2 m_A)$ as happens for the decay into taus. 

Once we include all $\Delta_{b,\tau}$ corrections as well as the decay into staus, the heavy Higgs branching ratio into taus is approximately given by
 \be
 \label{brstau}
{\rm{BR}}(H, A \to \tau^+ \tau^-) \propto  \left[\left( 3 \frac{m_b^2}{(1 + \Delta_b)^2 } +  \frac{M_W^2+M_Z^2}{\tan^2\beta}\right)\frac{ (1 + \Delta_\tau)^2}{m_{\tau}^2} + \left(1 + \frac{A_\tau^2}{m_A^2}\right) \right]^{-1}\, ,
\ee
where the term proportional to $(M_W^2+M_Z^2)$ is the approximate contribution from the decay into light charginos and neutralinos.  Similar to the case with heavy staus, Eq.~(\ref{brnostau}), the branching ratio is increased due to negative values of $\Delta_\tau$ and positive values of $\Delta_b$. However, comparing Eqs.~(\ref{brnostau}) and (\ref{brstau}), we see that this increase is partially compensated  for by the stau decays, quantified by the last term in  Eq.~(\ref{brstau}). Let us stress that Eq.~(\ref{brstau}) is only valid when the stau, chargino and neutralino masses are much smaller than $m_A$ and should be modified by the appropriate phase space factors if this is not the case.

As before, the production cross section is proportional to the product of the branching ratio times the bottom Yukawa squared, giving
\be
\label{tautaurate}
\sigma(pp\to (H,A)\to \tau^+ \tau^-) \propto \frac{ m_b^2 \tan^2\beta}{\left[\left(3 \frac{m_b^2}{m_{\tau}^2} + \frac{\left(M_W^2 +M_Z^2 \right) (1 + \Delta_b)^2}{m_\tau^2 \tan^2\beta} \right) (1+ \Delta_\tau)^2 + (1 + \Delta_b)^2\left(1 + \frac{A_\tau^2}{m_A^2}\right)\right]}\;.
\ee
The $\tau\tau$ production rate again increases due to negative $\Delta_\tau$ and decreases due to positive $\Delta_b$. However in addition, there is also a decrease in the rate due to the decays into the light staus.

\begin{figure}[tb]
\begin{center}
\includegraphics[width=0.45\textwidth]{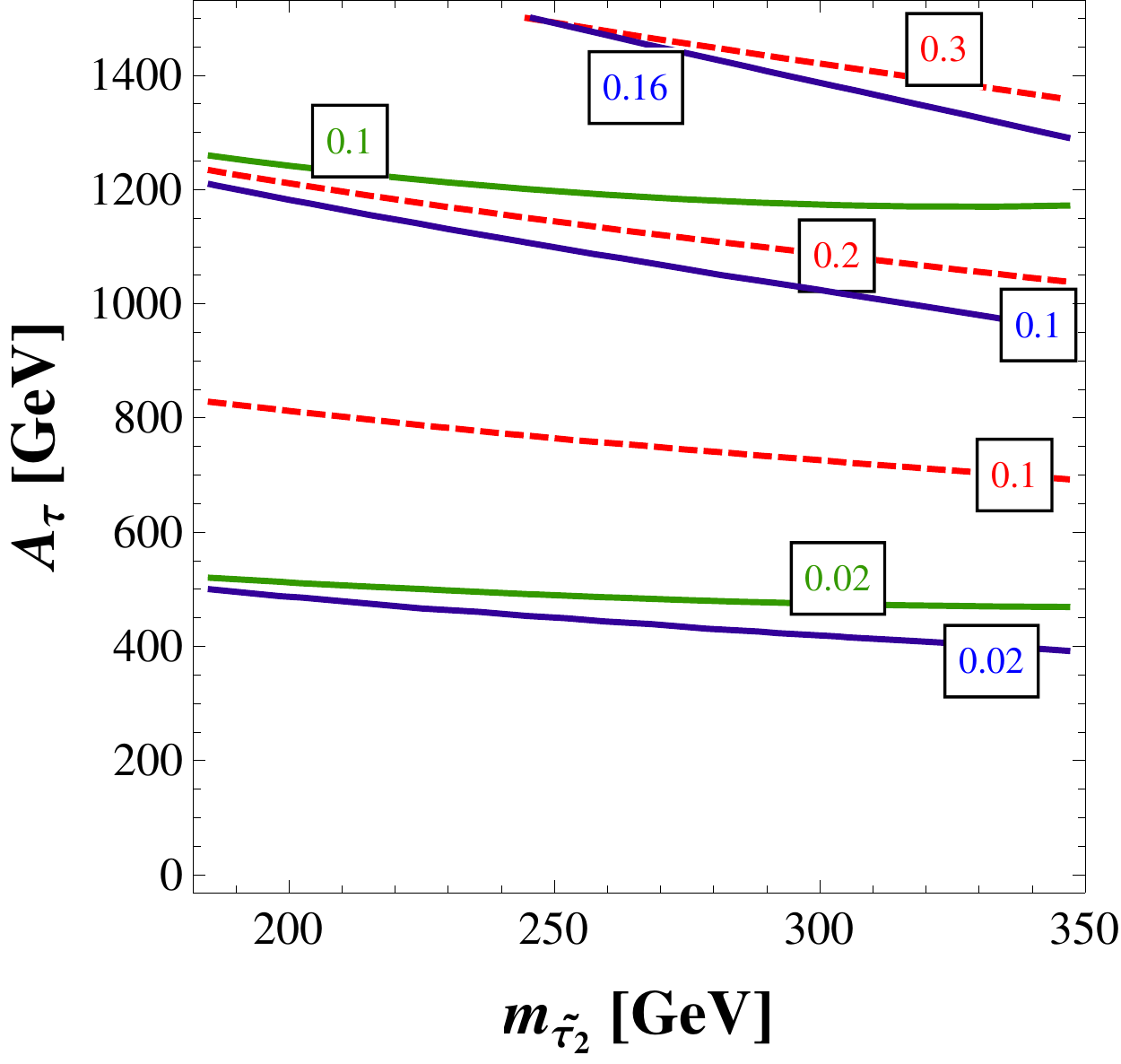} 
\includegraphics[width=0.45\textwidth]{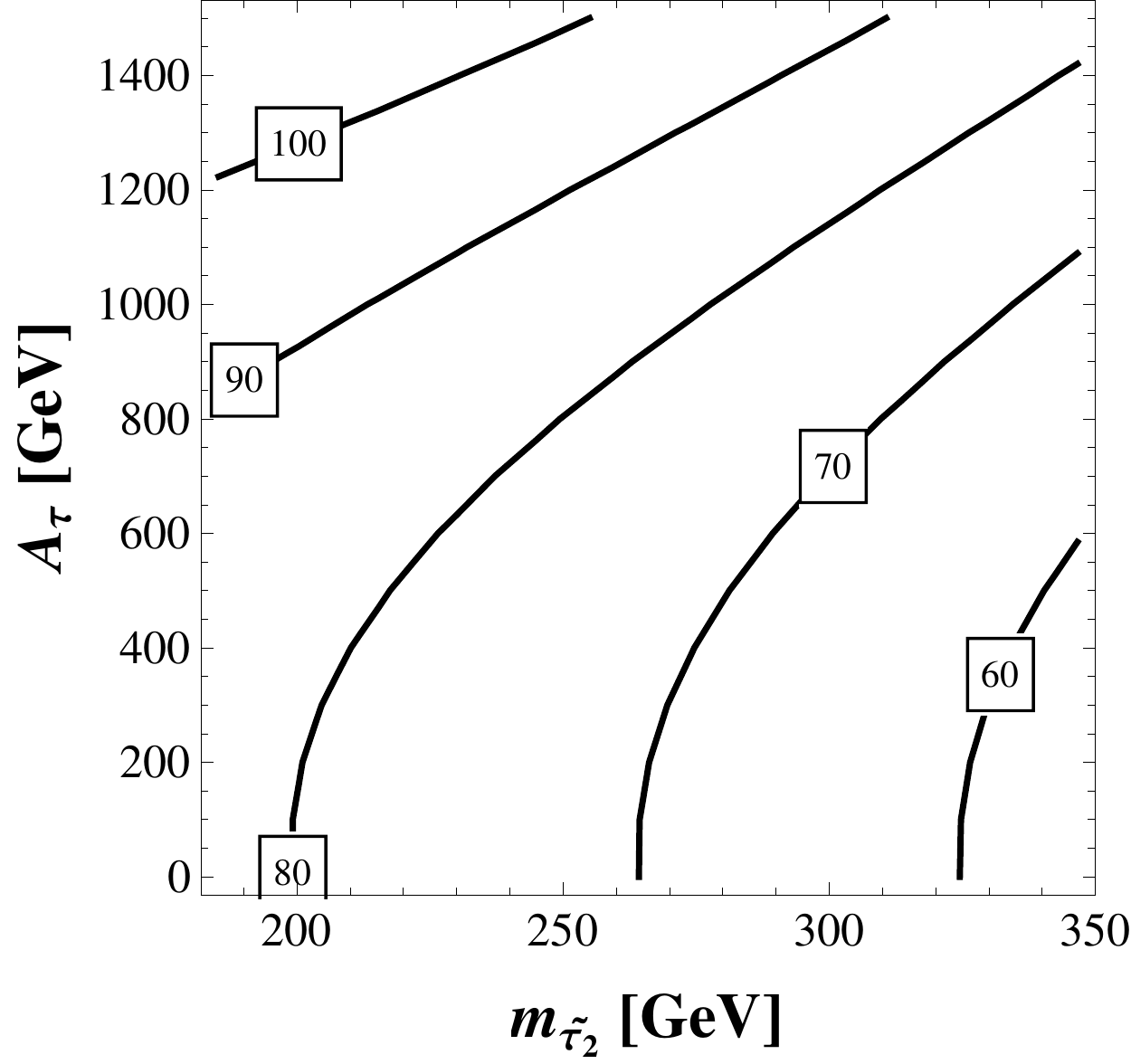} 
\end{center}
\caption{\textit{Left}: Branching ratios of the heavy Higgs bosons, $H$ and $A$. Dashed red lines: BR($A\to\tilde\tau_1\tilde\tau_2$), solid blue lines: BR($H\to\tilde\tau_1\tilde\tau_1$),  solid green lines: BR($H\to\tilde\tau_2\tilde\tau_2$). \textit{Right}: Total width of the heavy Higgs bosons in GeV. Mass of the lightest stau is fixed to 95 GeV and $m_A=1$ TeV.}
\label{mA1000CPEv}
\end{figure}

Let us now compare the $\tau$ branching ratio in the light stau scenario with the one that is obtained for heavy staus and small values of $\Delta_b \simeq 0.25$ and $\Delta_\tau \simeq 0$, as happens at $\tan\beta \simeq 70$ in the $m_h^{\rm max}$  scenario presented by ATLAS and CMS. For $A_\tau = m_A = 1$ TeV,  the ratio of the square of the running bottom and tau masses at the scale $m_A$ is approximately equal to 2. Therefore, from Eq.~(\ref{brstau}), for $\Delta_b = 0.25$, $\Delta_\tau =0$ and setting the $A_\tau$ term to 0 for heavy staus, the branching ratio of the decay of the heavy Higgs bosons into tau leptons is $\sim$~15\%. Considering instead a light stau, with $m_{\tilde{\tau}_1} \simeq 95$~GeV and stop masses giving  a 125 GeV Higgs for $m_{Q_3} = m_{u_3}=1$ TeV,  $m_{d_3}=800$~GeV, $M_3 = 1.2$ TeV, $\tan\beta =$ 70, $\Delta_\tau \simeq -0.15,\; \Delta_b \simeq 0.5$, the  branching ratio into taus is $\sim$ 20\%. Hence we see that in our scenario  there could be relevant modifications of the heavy Higgs decay branching ratio into $\tau$ pairs.

\begin{figure}[tb]
\begin{center}
\includegraphics[width=0.5\textwidth]{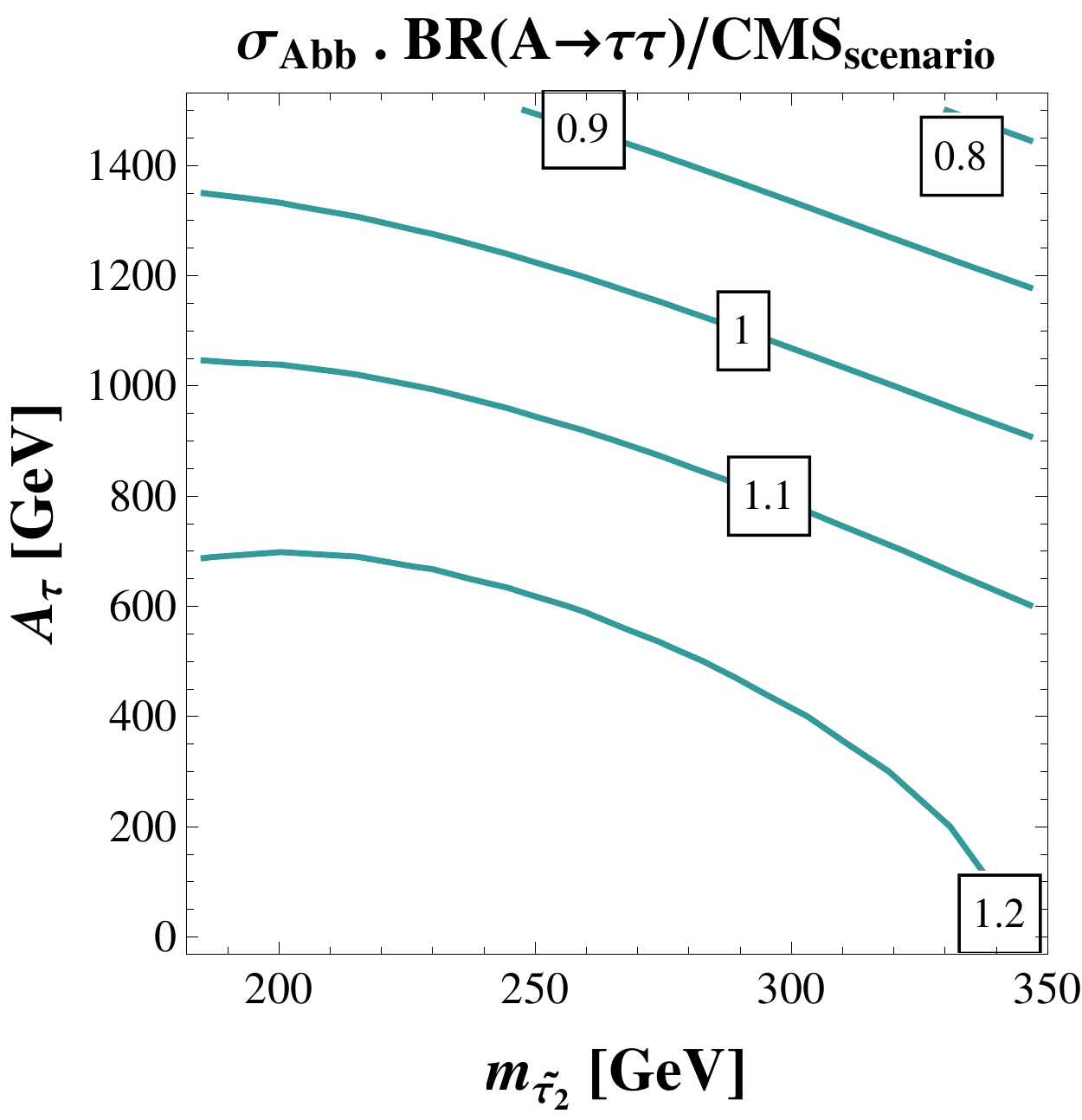} 
\end{center}
\caption{Production rate of  $\tau^+\tau^-$ induced by the presence of heavy CP-even and CP-odd scalars, with $m_A \simeq 1$~TeV, normalized to the rate obtained in the $m_h^{\rm max}$  scenario used by the CMS collaboration~\cite{LHCtautauMSSM}.}
\label{mAH_tb}
\end{figure}

Fig.~\ref{mA1000CPEv} shows the variation of the total width and the branching ratio into staus of the heavy CP-odd and CP-even Higgs bosons as a function of $m_{\tilde{\tau}_2}$ and $A_\tau$. We fix $\tan\beta = 70$, $m_{L_3} = m_{e_3}$, $M_1=40$ GeV, $m_A=1$ TeV and $m_{\tilde{\tau}_1} \simeq 95$~GeV. For a fixed value of  $m_{\tilde\tau_2}$  and increasing values of $A_\tau$, the figure in the left panel shows the expected increase of the decay rate into staus. The right panel shows the corresponding  increase in the total width with increasing $A_\tau$ and fixed $m_{\tilde \tau_2}$, which implies a  decrease of the branching ratio of the heavy Higgs decay into $\tau$ leptons.  On the other hand, for a fixed value of $A_\tau$, the value of  $\mu$ increases with $m_{\tilde{\tau}_2}$, which leads to an increase in $\Delta_b$ and a more negative $\Delta_\tau$.  Since the width of the decay into bottom quarks is the dominant one, the total width decreases.  However, note that negative $\Delta_\tau$ leads to an increase of the width of the decay into $\tau$ leptons,  and hence to an increase of the branching ratio of the decay of the heavy, non-standard Higgs bosons into these particles.  On the other hand, the production cross section of non-standard Higgs bosons is inversely proportional to $(1+ \Delta_b)^2$ and hence there is a compensating effect on the total rate of these Higgs bosons decaying into $\tau\tau$, Eq.~(\ref{tautaurate}).

Fig.~\ref{mAH_tb} shows the variation of the production rate of taus as a function of $m_{\tilde{\tau}_2}$ and $A_\tau$ with respect to the $m_h^{\rm max}$  scenario~\cite{Benchmark4} used by ATLAS and CMS~\cite{LHCtautauMSSM}. We use the same set of parameters as for Fig.~\ref{mA1000CPEv}. For a fixed value of $A_\tau$,  as a result of the compensation of effects discussed above,  only a small variation of the rate of $\tau\tau$ production is observed in the region of parameter space under analysis. On the other hand, for a given value of $m_{\tilde{\tau}_2}$  and increasing values of $A_\tau$, the $\tau\tau$ production rate decreases due to an increase of the width of the decay into stau. Therefore, only for large values of $A_\tau$ can we expect to significantly alleviate the experimental constraints on $m_A$ coming from the decay to taus. However, note that large values of $A_\tau > 1$~TeV lead to problems with vacuum stability in this region of parameter space~\cite{Stability}.

\section{Light Staus and Higgs Searches}
\label{Coll}

\begin{figure}[t]
\begin{center}
\begin{tabular}{ c}
\includegraphics[width=0.8\textwidth]{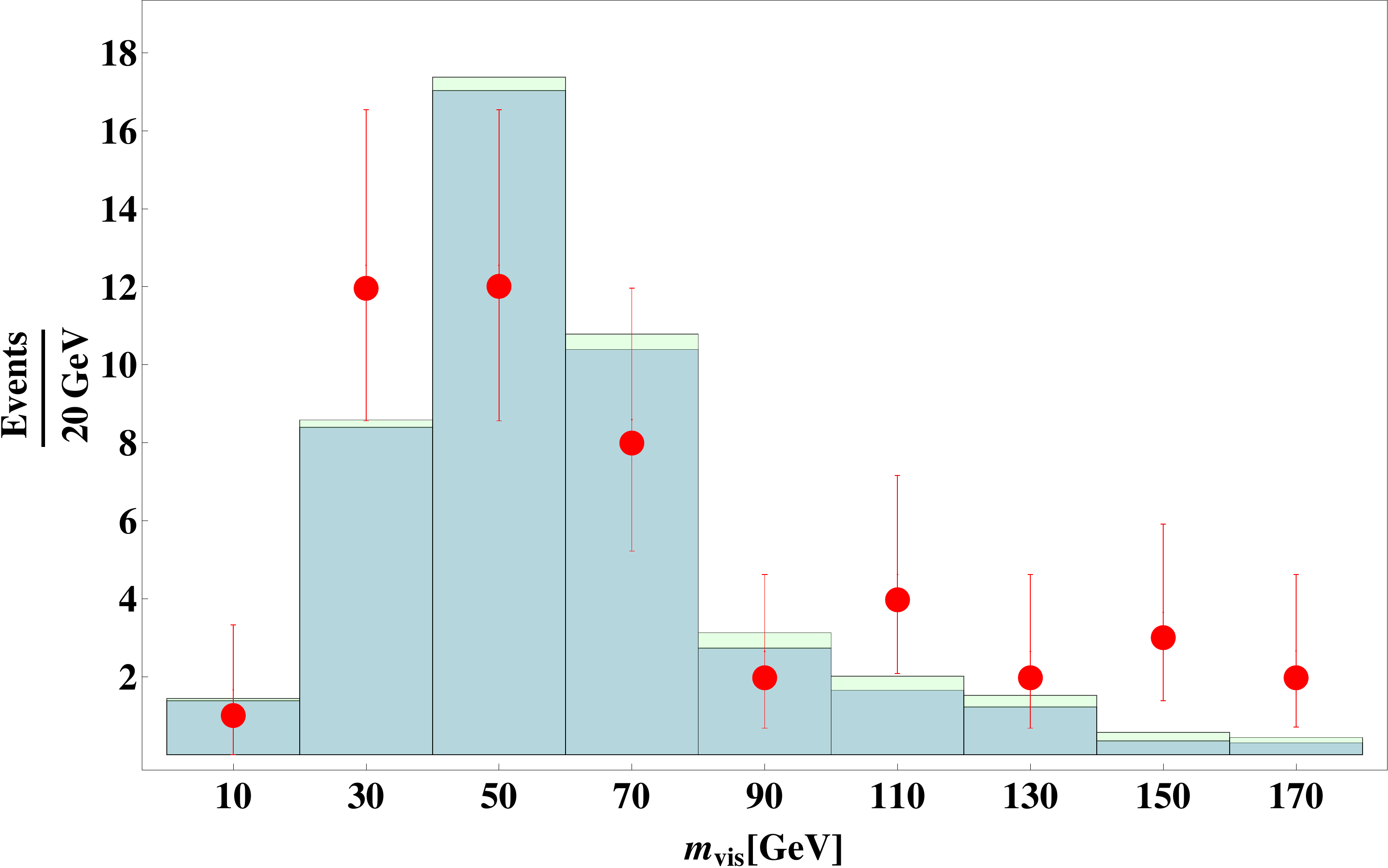} 
\end{tabular}
\end{center}
\caption{Visible mass distribution from signal $(pp \to \tilde{\nu}_\tau \tilde{\tau}_1 \to \ell\ell\tau_h+{\rm{MET}})$. The background distribution is taken from~\cite{cms_HZtautau} and is shown in blue. The red dots denote the CMS data points. }
\label{fig:tau_mvis}
\end{figure}

Light staus remain the distinctive signal of the MSSM scenario considered in this paper. In Ref.~\cite{Carena:2012gp}, we studied the possibility of searching for  them in the channel $(pp \to \tilde{\nu}_\tau \tilde{\tau}_1 \to W \tau \bar{\tau} + 2 \chi_0)$  at the LHC using a straight cut and count method. We specifically analyzed the final state signature consisting of one lepton, 2 hadronic taus and missing energy. We showed that this is a challenging search channel for both the 8 TeV and the 14 TeV runs, due to low statistics.

Here we will briefly mention another possibility  of probing our framework at the LHC. We note that the final state mentioned above is the same as the one arising in the Higgs search channel $(pp \to W h )$ followed by $(h \to \tau \bar \tau)$.  Therefore, it is interesting to see whether any present Higgs searches in this channel already have sensitivity to this new signal. 
 
Such searches typically require one hadronic tau and one leptonic tau. A common  variable used in these analyses~\cite{cms_HZtautau} is the visible mass, namely the invariant mass between the subleading light lepton and the hadronic tau. In Fig~\ref{fig:tau_mvis}, we present the visible mass distribution from our signal after imposing the main cuts presented in Ref.~\cite{cms_HZtautau}, namely $p_T^{\ell_1}>20$ GeV, $p_T^{\ell_2}>10$ GeV, $|\eta^{\tau_h}|<2.3$, either, $|\eta^{\mu}|<2.4$, or for the case of an electron $|\eta^{e}|<2.5$, and $L_T>80 $GeV, where $L_T$ is defined as the scalar sum of the $E_T$ of the lepton candidates. Parton level events are generated by {\texttt Madgraph5}~\cite{Alwall:2011uj} and taus are decayed using {\texttt Tauola}~\cite{Jadach:1990mz}. We note that the distribution peaks at larger values than both the distribution obtained from a Higgs with mass of about 125 GeV and the background distribution. Imposing an additional cut, $m_{\rm{vis}}>80$ GeV, would make the signal and background of the same order of magnitude. However, with the present amount of data (5 fb$^{-1}$ at the 7 TeV LHC and 12 fb$^{-1}$ at the 8 TeV LHC) the signal amounts to only $\sim 3$ events.  Therefore, similar to the case we analyzed before in Ref.~\cite{Carena:2012gp},  one would need  large statistics to claim the observation of light staus in these searches.

\section{Conclusions}
\label{Conc}

The LHC has recently discovered a bosonic resonance, with a mass close to 125~GeV and with properties similar to the SM Higgs particle. In this article we identify it with the lightest CP-even Higgs boson of the MSSM for large values of the CP-odd Higgs mass, $m_A$. Further, we study the possible modifications of the Higgs properties associated with the presence of light stops and light staus in the spectrum.

The most important effect of light staus is the possible modification of the Higgs diphoton decay width.  For values of $\tan\beta$ and of the stau mass parameters consistent with vacuum stability and with perturbativity up to energies of the order of the GUT scale, enhancements of the Higgs diphoton branching ratio of up to 50~\% with respect to the SM value are possible. Somewhat larger values may be obtained if one of these conditions is relaxed.  

Beyond light staus, light stops may also appear in the spectrum. We showed that stops lighter than about 200 GeV, while being consistent with a 125 GeV Higgs mass, can change the Higgs gluon fusion production rate as well as the Higgs diphoton branching ratio in a relevant way. There is a correlation between the NP effects in the Higgs diphoton branching ratio and the ones in the gluon fusion production rate: an enhancement in the diphoton branching ratio corresponds to a suppression in the Higgs production cross section and vice versa. The product of these effects tends to be governed by the behavior of the gluon fusion cross section. For instance, additional enhancements of about 30~\% of  the Higgs diphoton branching ratio may be obtained for sufficiently light stops, beyond the value obtained in the presence of only light staus.  However, in the same region of parameter space where these large enhancements take place, the gluon fusion production rate is strongly suppressed, leading to an overall suppression of the diphoton production rate in gluon fusion processes. As a result, this diphoton production rate may be enhanced by at most an additional 10~\% in the presence of light stops, but could also be suppressed by even larger amounts. Moreover, we showed that a discrepancy between the diphoton production rates  from Higgs produced in gluon fusion and Higgs produced in weak boson fusion processes, normalized to their SM values,  may be a clear indication of light stops in the spectrum.   

Light stop masses are highly constrained by  both the Tevatron and the LHC experiments. Dedicated stop searches show that if the lightest neutralino is lighter than $\sim 100$ GeV, the only stop masses below $\sim 250$ GeV not in conflict with experiments are those close to the threshold  for stop decays into a top and a neutralino. However,  we showed that in the presence of light staus a new decay channel opens up, $(\tilde t\to\tilde \tau^+_1 b\nu)$, allowing stops to evade current experimental bounds in the range 120-140 GeV.  

Further, light staus may induce relevant CP-even Higgs mixing effects. These can lead to a modification of  the $(h \to \tau^+\tau^-)$ decay branching ratio by relevant amounts while inducing  only a small modification of the dominant $(h\to b\bar{b})$ decay branching ratio. However, the stau mass parameters leading to these effects are very strongly constrained by the requirement of vacuum stability and perturbativity up to the GUT scale. Imposing these two conditions, variations of only a few percent due to NP effects can be obtained for the ratio of BR$(h\to b\bar{b})$ to BR$(h\to \tau\bar{\tau})$. However, we also show that larger modification can be achieved if we relax the vacuum stability or perturbativity constraints.  We present such a case as an example and show that a suppression of the branching ratio of the Higgs decay into a pair of $\tau$s of about 20 $\%$ may be induced by these mixing effects, while BR$(h\to b\bar{b})$ remains SM-like.

Relevant Higgs mixing effects are associated with  large values of the stau mixing parameter, $A_\tau$, and moderate values of $m_A$. Therefore, decays of the heavy Higgs bosons into staus may become important.  
Additionally, in our scenario, sizable threshold corrections to the $\tau$ and bottom masses arise. We showed that both these effects together lead to relevant modifications of the width of the heavy Higgs decay into $\tau$ leptons compared to the ones in the $m_h^{\rm max}$  scenario analyzed by the LHC experiments.  This in turn may relax the experimental bounds on $m_A$ for large values of $\tan \beta$ in the presence of a large stau mixing parameter, $A_\tau$.

Finally, we point out an interesting possibility to test light staus at the LHC. We stress that the final state arising from the associated production of Higgs bosons with $W$ bosons, with the Higgs decaying into $\tau$ leptons, is the same as in the associated production of staus with sneutrinos.  We therefore consider using existing Higgs searches to put constraints on the production of light staus. We conclude that the stau-induced production rates are currently too small to put relevant constraints in this scenario. However, these type of searches could be useful  to probe our scenario in the future.

In summary, we have shown that, beyond the enhancement of the lightest CP-even Higgs diphoton decay rate, the Higgs and supersymmetric particle phenomenology associated with light stops and staus in the spectrum is quite rich and may be explored at the LHC in the coming years. 

\vspace{1cm}

{\bf ACKNOWLEDGEMENTS:}
We would like to thank Felix Yu for useful discussions. SG thanks the Galileo Galilei Institute for Theoretical Physics for its hospitality during some part of this work. L.T.W. is supported by the NSF under grant PHY-0756966 and the DOE
Early Career Award under grant de-sc0003930. Fermilab is operated by Fermi Research Alliance, LLC under Contract No. DE-AC02-07CH11359 with the U.S. Department of Energy. Work at ANL is supported in part by the U.S. Department of Energy under Contract No. DE-AC02-06CH11357. N.R.S is supported by the DoE grant No. DE-SC0007859.  We would also like to thank the Aspen Center for Physics and the  KITP, Santa Barbara,  where part of the work has been done.

%


\begin{thebibliography}{999}
\bibitem{:2012gk} 
  G.~Aad {\it et al.}  [ATLAS Collaboration],
  Phys.\ Lett.\ B {\bf 716}, 1 (2012)
  [arXiv:1207.7214 [hep-ex]].

\bibitem{:2012gu} 
  S.~Chatrchyan {\it et al.}  [CMS Collaboration],
  Phys.\ Lett.\ B {\bf 716}, 30 (2012)
  [arXiv:1207.7235 [hep-ex]].
  
  \bibitem{Diphoton}   ATLAS Collaboration,   ATLAS-CONF-2013-012
  
\bibitem{Carena:2011aa} 
  M.~Carena, S.~Gori, N.~R.~Shah and C.~E.~M.~Wagner,
  JHEP {\bf 1203}, 014 (2012)
  [arXiv:1112.3336 [hep-ph]].

\bibitem{Carena:2012gp} 
  M.~Carena, S.~Gori, N.~R.~Shah, C.~E.~M.~Wagner and L.~-T.~Wang,
  JHEP {\bf 1207}, 175 (2012)
  [arXiv:1205.5842 [hep-ph]].

\bibitem{Blum:2012ii} 
  K.~Blum, R.~T.~D'Agnolo and J.~Fan,
  JHEP {\bf 1301}, 057 (2013)
  [arXiv:1206.5303 [hep-ph]].
  
\bibitem{SchmidtHoberg:2012yy} 
  K.~Schmidt-Hoberg and F.~Staub,
  JHEP {\bf 1210}, 195 (2012)
  [arXiv:1208.1683 [hep-ph]].
 
\bibitem{Kniehl:1995tn} 
  B.~A.~Kniehl and M.~Spira,
  Z.\ Phys.\ C {\bf 69}, 77 (1995)
  [hep-ph/9505225].
  
\bibitem{CLW}
    M.~Carena, I.~Low and C.~E.~M.~Wagner,
  JHEP {\bf 1208}, 060 (2012)
  [arXiv:1206.1082 [hep-ph]].
  
  
\bibitem{Batell:2011pz} 
  B.~Batell, S.~Gori and L.~-T.~Wang,
  JHEP {\bf 1206}, 172 (2012)
  [arXiv:1112.5180 [hep-ph]];
 J.~Cao, Z.~Heng, J.~M.~Yang, Y.~Zhang and J.~Zhu,
  JHEP {\bf 1203}, 086 (2012)
  [arXiv:1202.5821 [hep-ph]];
  A.~Akeroyd and S.~Moretti, 
Phys.\ Rev.\ {\bf D86} (2012) 035015 [arXiv:1206.0535 [hep-ph]];
  W.-F. Chang, J.~N. Ng, and J.~M. Wu, 
  Phys.\ Rev.\ {\bf D86} (2012) 033003 [arXiv:1206.5047 [hep-ph]];
 H.~An, T.~Liu and L.~-T.~Wang,
  Phys.\ Rev.\ D {\bf 86}, 075030 (2012)
  [arXiv:1207.2473 [hep-ph]];
  A.~Joglekar, P.~Schwaller and C.~E.~M.~Wagner,
  arXiv:1207.4235 [hep-ph];
  N.~Arkani-Hamed, K.~Blum, R.~T.~D'Agnolo and J.~Fan,
  arXiv:1207.4482 [hep-ph];
  N.~Haba, K.~Kaneta, Y.~Mimura, and R.~Takahashi,
 arXiv:1207.5102 [hep-ph];
   L.~G.~Almeida, E.~Bertuzzo, P.~A.~N.~Machado and R.~Z.~Funchal,
  arXiv:1207.5254 [hep-ph];
%
  G.~F.~Giudice, P.~Paradisi, A.~Strumia and A.~Strumia,
  JHEP {\bf 1210}, 186 (2012)
  [arXiv:1207.6393 [hep-ph]].
  A.~Delgado, G.~Nardini and M.~Quiros,
  Phys.\ Rev.\ D {\bf 86}, 115010 (2012)
  [arXiv:1207.6596 [hep-ph]].
%
R.~Sato, K.~Tobioka, and N.~Yokozaki
 Phys.\ Lett.\   {\bf B716} (2012) 441--445 [arXiv:1208.2630 [hep-ph]];
%
H.~Davoudiasl, H.-S. Lee, and W.~J. Marciano, 
 arXiv:1208.2973 [hep-ph];
%
M.~Voloshin, 
 arXiv:1208.4303 [hep-ph];
  D.~McKeen, M.~Pospelov and A.~Ritz,
  Phys.\ Rev.\ D {\bf 86}, 113004 (2012)
  [arXiv:1208.4597 [hep-ph]].
   A.~Urbano, 
 arXiv:1208.5782 [hep-ph];
%
E.~J. Chun, H.~M. Lee, and P.~Sharma, 
 arXiv:1209.1303 [hep-ph];
%
%
L.~Wang and X.-F. Han, 
   arXiv:1209.0376 [hep-ph];
%
%
 M.~Chala,
  arXiv:1210.6208 [hep-ph];
%
%
 H.~Davoudiasl, I.~Lewis and E.~Ponton,
  arXiv:1211.3449 [hep-ph].
      B.~Batell, S.~Jung and H.~M.~Lee,
  arXiv:1211.2449 [hep-ph];
  J.~Kearney, A.~Pierce and N.~Weiner,
  Phys.\ Rev.\ D {\bf 86}, 113005 (2012)
  [arXiv:1207.7062 [hep-ph]];
  R.~Huo, G.~Lee, A.~M.~Thalapillil and C.~E.~M.~Wagner,
  arXiv:1212.0560 [hep-ph];
  A.~Joglekar, P.~Schwaller and C.~E.~M.~Wagner,
  arXiv:1303.2969 [hep-ph].
  
\bibitem{Hall:2011aa}
  L.~J.~Hall, D.~Pinner and J.~T.~Ruderman,
  arXiv:1112.2703 [hep-ph].
  U.~Ellwanger,
  JHEP {\bf 1203}, 044 (2012)
  [arXiv:1112.3548 [hep-ph]].
  G.~Belanger, U.~Ellwanger, J.~F.~Gunion, Y.~Jiang, S.~Kraml and J.~H.~Schwarz,
  JHEP {\bf 1301}, 069 (2013)
  [arXiv:1210.1976 [hep-ph]].
  
\bibitem{Batell:2012mj} 
  B.~Batell, D.~McKeen and M.~Pospelov,
  JHEP {\bf 1210}, 104 (2012)
  [arXiv:1207.6252 [hep-ph]].
  C.~Cheung, S.~D.~McDermott and K.~M.~Zurek,
  arXiv:1302.0314 [hep-ph].
  
  \bibitem{Dibosons} CMS Collaboration, CMS-PAS-HIG-13-002; CMS-PAS-HIG-13-003; \\
  ATLAS Collaboration, ATLAS-CONF-2013-013; ATLAS-CONF-2012-158.
  
    \bibitem{CMSdiphoton} CMS-HIG-13-001.
  
  
  \bibitem{LHCtautau} 
ÊG.~Aad {\it et al.} Ê[ATLAS Collaboration], Ê
Phys.\ Lett.\ B {\bf 705}, 174 (2011) Ê[arXiv:1107.5003 [hep-ex]];
S.~Chatrchyan {\it et al.} Ê[CMS Collaboration], 
Ê
ÊarXiv:1202.4083 [hep-ex]; Ê
  Atlas Collaboration, ATLAS-CONF-2012-160; CMS Collaboration, CMS-PAS-HIG-12-043.
  
\bibitem{Aaltonen:2012qt} 
  T.~Aaltonen {\it et al.}  [CDF and D0 Collaborations],
  Phys.\ Rev.\ Lett.\  {\bf 109}, 071804 (2012)
  [arXiv:1207.6436 [hep-ex]].
  
\bibitem{Chatrchyan:2012ww} 
  S.~Chatrchyan {\it et al.}  [CMS Collaboration],
  Phys.\ Lett.\ B {\bf 710}, 284 (2012)
  [arXiv:1202.4195 [hep-ex]].
  
    
\bibitem{Aad:2012gxa} 
  G.~Aad {\it et al.}  [ATLAS Collaboration],
  Phys.\ Lett.\ B {\bf 718}, 369 (2012)
  [arXiv:1207.0210 [hep-ex]].
  
      
\bibitem{Dobrescu:2012td} 
  B.~A.~Dobrescu and J.~D.~Lykken,
  JHEP {\bf 1302}, 073 (2013)
  [arXiv:1210.3342 [hep-ph]].
  
     \bibitem{cms_HZtautau}
  [CMS Collaboration], CMS PAS HIG-12-051
  
\bibitem{Ellis:1975ap} 
  J.~R.~Ellis, M.~K.~Gaillard and D.~V.~Nanopoulos,
  Nucl.\ Phys.\ B {\bf 106}, 292 (1976).
  
\bibitem{Shifman:1979eb} 
  M.~A.~Shifman, A.~I.~Vainshtein, M.~B.~Voloshin and V.~I.~Zakharov,
  Sov.\ J.\ Nucl.\ Phys.\  {\bf 30}, 711 (1979)
  [Yad.\ Fiz.\  {\bf 30}, 1368 (1979)].
  
\bibitem{Diaz:2004qt} 
  M.~A.~Diaz and P.~Fileviez Perez,
  J.\ Phys.\ G {\bf 31}, 563 (2005)
  [hep-ph/0412066].
  
  
\bibitem{Altmannshofer:2012ar} 
  W.~Altmannshofer, S.~Gori and G.~D.~Kribs,
  Phys.\ Rev.\ D {\bf 86}, 115009 (2012)
  [arXiv:1210.2465 [hep-ph]].

  
  
  
\bibitem{Ratz:2008qh} 
  M.~Ratz, K.~Schmidt-Hoberg and M.~W.~Winkler,
  JCAP {\bf 0810}, 026 (2008)
  [arXiv:0808.0829 [hep-ph]].
  
  
\bibitem{Hisano:2010re} 
  J.~Hisano and S.~Sugiyama,
  Phys.\ Lett.\ B {\bf 696}, 92 (2011)
  [arXiv:1011.0260 [hep-ph]].
 
 \bibitem{Stability}
 M.~Carena, S.~Gori, I.~Low, N.~R.~Shah and C.~E.~M.~Wagner,
  JHEP {\bf 1302}, 114 (2013)
  [arXiv:1211.6136 [hep-ph]].
  
\bibitem{Kitahara:2012pb} 
  T.~Kitahara,
  JHEP {\bf 1211}, 021 (2012)
  [arXiv:1208.4792 [hep-ph]].
  
 \bibitem{deltamb}
   L.~J.~Hall, R.~Rattazzi and U.~Sarid,
  Phys.\ Rev.\ D {\bf 50}, 7048 (1994)
  [hep-ph/9306309];
 R.~Hempfling,
  Phys.\ Rev.\ D {\bf 49}, 6168 (1994).

\bibitem{deltamb1} 
 M.~S.~Carena, M.~Olechowski, S.~Pokorski and C.~E.~M.~Wagner,
  Nucl.\ Phys.\ B {\bf 426}, 269 (1994)
  [hep-ph/9402253].
 
\bibitem{deltamb2}   D.~M.~Pierce, J.~A.~Bagger, K.~T.~Matchev and R.~-j.~Zhang,
  Nucl.\ Phys.\ B {\bf 491}, 3 (1997)
  [hep-ph/9606211].

\bibitem{deltamb3}   J.~Guasch, W.~Hollik and S.~Penaranda,
  Phys.\ Lett.\ B {\bf 515}, 367 (2001)
  [hep-ph/0106027];
   H.~E.~Haber, M.~J.~Herrero, H.~E.~Logan, S.~Penaranda, S.~Rigolin and D.~Temes,
  Phys.\ Rev.\ D {\bf 63}, 055004 (2001)
  [hep-ph/0007006].



  \bibitem{LEPstau} LEP2 SUSY Working Group, LEPSUSYWG/04-01.1;  \\
    A.~Heister {\it et al.}  [ALEPH Collaboration],
  Phys.\ Lett.\ B {\bf 526}, 206 (2002)
  [hep-ex/0112011]; \\
    J.~Abdallah {\it et al.}  [DELPHI Collaboration],
  Eur.\ Phys.\ J.\ C {\bf 31}, 421 (2003)
  [hep-ex/0311019]; \\
 P.~Achard {\it et al.}  [L3 Collaboration],
  Phys.\ Lett.\ B {\bf 580}, 37 (2004)
  [hep-ex/0310007]; \\
   G.~Abbiendi {\it et al.}  [OPAL Collaboration],
  Eur.\ Phys.\ J.\ C {\bf 32}, 453 (2004)
  [hep-ex/0309014].

  
   
\bibitem{Kitahara:2013lfa} 
  T.~Kitahara and T.~Yoshinaga,
  arXiv:1303.0461 [hep-ph].
  
  \bibitem{Carena:1999bh}
  M.~S.~Carena, S.~Mrenna and C.~E.~M.~Wagner,
 Phys.\ Rev.\  D {\bf 62}, 055008 (2000)
  [arXiv:hep-ph/9907422];
%
  M.~S.~Carena, S.~Mrenna and C.~E.~M.~Wagner,
  Phys.\ Rev.\ D\ {\bf 60}, 075010  (1999)
  [hep-ph/9808312];
 M.~S.~Carena, H.~E.~Haber, H.~E.~Logan and S.~Mrenna,
  Phys.\ Rev.\ D {\bf 65}, 055005 (2002)
  [Erratum-ibid.\ D {\bf 65}, 099902 (2002)]
  [hep-ph/0106116].



\bibitem{Haisch:2012re} 
  U.~Haisch and F.~Mahmoudi,
  JHEP {\bf 1301}, 061 (2013)
  [arXiv:1210.7806 [hep-ph]].
  
\bibitem{Altmannshofer:2012ks} 
  W.~Altmannshofer, M.~Carena, N.~R.~Shah and F.~Yu,
  JHEP {\bf 1301}, 160 (2013)
  [arXiv:1211.1976 [hep-ph]].
  



\bibitem{Djouadi:1998az}
  A.~Djouadi,
  Phys.\ Lett.\ B {\bf 435}, 101 (1998)
  [hep-ph/9806315].
  
\bibitem{Dermisek:2007fi} 
  R.~Dermisek and I.~Low,
  Phys.\ Rev.\ D {\bf 77}, 035012 (2008)
  [hep-ph/0701235 [HEP-PH]].
  


\bibitem{Sbottomlimits} ATLAS Collaboration, ATLAS-CONF-2012-165; CMS Collaboration, PAS-SUS-12-028.


\bibitem{Okada:1990vk}
  Y.~Okada, M.~Yamaguchi and T.~Yanagida,
  Prog.\ Theor.\ Phys.\  {\bf 85}, 1 (1991).

\bibitem{Ellis:1990nz}
  J.~R.~Ellis, G.~Ridolfi and F.~Zwirner,
  Phys.\ Lett.\  B {\bf 257}, 83 (1991).

\bibitem{Haber:1990aw}
  H.~E.~Haber and R.~Hempfling,
  Phys.\ Rev.\ Lett.\  {\bf 66}, 1815 (1991).

\bibitem{mhiggsRG1a}
  J.~A.~Casas, J.~R.~Espinosa, M.~Quiros and A.~Riotto,
  Nucl.\ Phys.\  B {\bf 436}, 3 (1995)
  [Erratum-ibid.\  B {\bf 439}, 466 (1995)]
  [arXiv:hep-ph/9407389].

\bibitem{mhiggsRG1} M.~Carena, J.~Espinosa, M.~Quir\'os and C.~Wagner,
                    {\em Phys. Lett.} {\bf B 355} (1995) 209,
                    hep-ph/9504316;\\
                    M.~Carena, M.~Quir\'os and C.~Wagner,
                    {\em Nucl. Phys.} {\bf B 461} (1996) 407,
                    hep-ph/9508343.
\bibitem{HHH} H.~Haber, R.~Hempfling and A.~Hoang,
              {\em Z. Phys.} {\bf C 75} (1997) 539,
              hep-ph/9609331.

\bibitem{Heinemeyer:1998yj}
  S.~Heinemeyer, W.~Hollik and G.~Weiglein,
  Comput.\ Phys.\ Commun.\  {\bf 124}, 76 (2000)
  [hep-ph/9812320].
\bibitem{Heinemeyer:1998np}
  S.~Heinemeyer, W.~Hollik and G.~Weiglein,
  Eur.\ Phys.\ J.\ C {\bf 9}, 343 (1999)
  [hep-ph/9812472].
  \bibitem{Carena:2000dp}
  M.~S.~Carena, H.~E.~Haber, S.~Heinemeyer, W.~Hollik, C.~E.~M.~Wagner and G.~Weiglein,
  Nucl.\ Phys.\ B {\bf 580}, 29 (2000)
  [hep-ph/0001002].

\bibitem{Martin:2002wn}
  S.~P.~Martin,
  Phys.\ Rev.\ D {\bf 67}, 095012 (2003)
  [hep-ph/0211366].

\bibitem{Degrassi:2002fi}
  G.~Degrassi, S.~Heinemeyer, W.~Hollik, P.~Slavich and G.~Weiglein,
  Eur.\ Phys.\ J.\ C {\bf 28}, 133 (2003)
  [hep-ph/0212020].

\bibitem{Arbey:2011ab}
  A.~Arbey, M.~Battaglia, A.~Djouadi, F.~Mahmoudi and J.~Quevillon,
  Phys.\ Lett.\  B {\bf 708}, 162 (2012)
  [arXiv:1112.3028 [hep-ph]].

\bibitem{Heinemeyer:2011aa}
  S.~Heinemeyer, O.~Stal and G.~Weiglein,
  Phys.\ Lett.\  B {\bf 710}, 201 (2012)
  [arXiv:1112.3026 [hep-ph]].

\bibitem{Draper:2011aa}
  P.~Draper, P.~Meade, M.~Reece and D.~Shih,
  arXiv:1112.3068 [hep-ph].



\bibitem{Benchmark4}  M.~Carena, S.~Heinemeyer, O.~Stal, C.~E.~M.~Wagner and G.~Weiglein,
  arXiv:1302.7033 [hep-ph]; 
  M.~S.~Carena, S.~Heinemeyer, C.~E.~M.~Wagner and G.~Weiglein,
  Eur.\ Phys.\ J.\ C {\bf 45}, 797 (2006)
  [hep-ph/0511023].

 
 \bibitem{Stopsglueglue} 
  K.~Blum, R.~D'Agnolo and J.~Fan,
  {\em JHEP} {\bf 1301} (2013) 057
  [arXiv:1206.5303 [hep-ph]]; \\
  M.~Buckley and D.~Hooper,
  {\em Phys.\ Rev.} {\bf D 86} (2012) 075008
  [arXiv:1207.1445 [hep-ph]];\\
  J.~Espinosa, C.~Grojean, V.~Sanz and M.~Trott,
  {\em JHEP} {\bf 1212} (2012) 077
  [arXiv:1207.7355 [hep-ph]].

\bibitem{CPsuperH}
  J.~S.~Lee, A.~Pilaftsis, M.~S.~Carena, S.~Y.~Choi, M.~Drees, J.~R.~Ellis and C.~E.~M.~Wagner,
  Comput.\ Phys.\ Commun.\  {\bf 156}, 283 (2004)
  [hep-ph/0307377]; \\
 J.~S.~Lee, M.~Carena, J.~Ellis, A.~Pilaftsis and C.~E.~M.~Wagner,
  Comput.\ Phys.\ Commun.\  {\bf 180}, 312 (2009)
  [arXiv:0712.2360 [hep-ph]].
J.~S.~Lee, M.~Carena, J.~Ellis, A.~Pilaftsis and C.~E.~M.~Wagner,
  Comput.\ Phys.\ Commun.\  {\bf 184}, 1220 (2013)
  [arXiv:1208.2212 [hep-ph]].
  
\bibitem{Drees:2012dd} 
  M.~Drees, M.~Hanussek and J.~S.~Kim,
  Phys.\ Rev.\ D {\bf 86}, 035024 (2012)
  [arXiv:1201.5714 [hep-ph]].
  
\bibitem{Alves:2012ft} 
  D.~S.~M.~Alves, M.~R.~Buckley, P.~J.~Fox, J.~D.~Lykken and C.~-T.~Yu,
  arXiv:1205.5805 [hep-ph].
  
\bibitem{Han:2012fw} 
  Z.~Han, A.~Katz, D.~Krohn and M.~Reece,
  JHEP {\bf 1208}, 083 (2012)
  [arXiv:1205.5808 [hep-ph]].
 
\bibitem{Kilic:2012kw} 
  C.~Kilic and B.~Tweedie,
  arXiv:1211.6106 [hep-ph].


  \bibitem{NausheenKaty} K. Freese and N.R. Shah, in preparation.

\bibitem{Delgado:2012eu} 
  A.~Delgado, G.~F.~Giudice, G.~Isidori, M.~Pierini and A.~Strumia,
  arXiv:1212.6847 [hep-ph].

\bibitem{Hikasa:1987db} 
  K.~-i.~Hikasa and M.~Kobayashi,
  Phys.\ Rev.\ D {\bf 36}, 724 (1987).
  
  C.~Boehm, A.~Djouadi and Y.~Mambrini,
  Phys.\ Rev.\ D {\bf 61}, 095006 (2000)
  [hep-ph/9907428].

\bibitem{Alwall:2011uj} 
  J.~Alwall, M.~Herquet, F.~Maltoni, O.~Mattelaer and T.~Stelzer,
  JHEP {\bf 1106}, 128 (2011)
  [arXiv:1106.0522 [hep-ph]].
  
\bibitem{Aaltonen:2012tq} 
  T.~Aaltonen {\it et al.}  [CDF Collaboration],
  JHEP {\bf 1210}, 158 (2012)
  [arXiv:1203.4171 [hep-ex]].
\bibitem{Kats:2011it} 
  Y.~Kats and D.~Shih,
  JHEP {\bf 1108}, 049 (2011)
  [arXiv:1106.0030 [hep-ph]].
  
\bibitem{Krizka:2012ah} 
  K.~Krizka, A.~Kumar and D.~E.~Morrissey,
  arXiv:1212.4856 [hep-ph].
  

  
\bibitem{Yu:2012kj} 
  Z.~-H.~Yu, X.~-J.~Bi, Q.~-S.~Yan and P.~-F.~Yin,
  arXiv:1211.2997 [hep-ph].
  
\bibitem{Abazov:2010pa} 
  V.~M.~Abazov {\it et al.}  [D0 Collaboration],
  Phys.\ Rev.\ D {\bf 82}, 071102 (2010)
  [arXiv:1008.4284 [hep-ex]].
\bibitem{cms11004}
 [CMS Collaboration], CMS TOP-11-004
\bibitem{Collaboration:2012hz} 
  T.~Aaltonen {\it et al.}  [CDF Collaboration],
  Phys.\ Rev.\ Lett.\  {\bf 109}, 192001 (2012)
  [arXiv:1208.5720 [hep-ex]].
\bibitem{Aad:2012vna} 
  G.~Aad {\it et al.}  [ATLAS Collaboration],
  arXiv:1211.7205 [hep-ex].
\bibitem{:2012cj} 
  S.~Chatrchyan {\it et al.}  [CMS Collaboration],
  arXiv:1212.6682 [hep-ex].



\bibitem{:2012bta} 
  S.~Chatrchyan {\it et al.}  [CMS Collaboration],
  JHEP {\bf 1211}, 067 (2012)
  [arXiv:1208.2671 [hep-ex]].
\bibitem{ATLAS:2012aa} 
  G.~Aad {\it et al.}  [ATLAS Collaboration],
  JHEP {\bf 1205}, 059 (2012)
  [arXiv:1202.4892 [hep-ex]].
  \bibitem{cmsllttbar}
 [CMS Collaboration], CMS TOP-12-007

  
  
  
\bibitem{:2012xh} 
  G.~Aad {\it et al.}  [ATLAS Collaboration],
  Phys.\ Lett.\ B {\bf 717}, 89 (2012)
  [arXiv:1205.2067 [hep-ex]].
  
\bibitem{Chatrchyan:2012vs} 
  S.~Chatrchyan {\it et al.}  [CMS Collaboration],
  Phys.\ Rev.\ D {\bf 85}, 112007 (2012)
  [arXiv:1203.6810 [hep-ex]].
    
  
\bibitem{Choudhury:2012kn} 
  A.~Choudhury and A.~Datta,
  Mod.\ Phys.\ Lett.\ A {\bf 27}, 1250188 (2012)
  [arXiv:1207.1846 [hep-ph]].
  
   
        \bibitem{ATLAS:166}
 ATLAS collaboration,
  ATLAS-CONF-2012-166.
  
          \bibitem{ATLAS:167}
 ATLAS collaboration,
  ATLAS-CONF-2012-167.
  
          \bibitem{ATLAS:001}
 ATLAS collaboration,
  ATLAS-CONF-2013-001.
  
          \bibitem{CMS:stopbchargino}
 CMS collaboration,
  CMS PAS SUS-12-023.
  
  
  \bibitem{LHCtautauMSSM} Atlas Collaboration, ATLAS-CONF-2012-094; CMS Collaboration, CMS-PAS-HIG-12-050.  
  
\bibitem{Jadach:1990mz} 
  S.~Jadach, J.~H.~Kuhn and Z.~Was,
  Comput.\ Phys.\ Commun.\  {\bf 64}, 275 (1990).


  
  
  
  
  
  



\end{thebibliography}
\end{document}